\documentclass[review]{elsarticle}
\pdfoutput=1
\usepackage[hidelinks,hyperfootnotes=false, bookmarks]{hyperref}
\usepackage{graphicx} 
\usepackage{mathtools}
\usepackage{amsmath}
\usepackage{amssymb}
\usepackage{algorithm}
\usepackage{upgreek}
\usepackage{algorithmic}
\usepackage[utf8]{inputenc}
\usepackage{multirow}
\usepackage{color,soul}
\usepackage{array}
\usepackage{graphicx}
\usepackage{epstopdf}
\usepackage[figurename=Fig.]{caption}
\usepackage{subcaption}
\usepackage{array}
\usepackage{amsfonts}
\usepackage{scalerel}
\usepackage{subcaption} 
\usepackage[labelformat=simple]{subcaption}
\usepackage[scientific-notation=true]{siunitx}

\DeclareMathOperator{\heavi}{\mathcal{H}}
\DeclareMathOperator{\tr}{tr}

\newcolumntype{M}[1]{>{\centering\arraybackslash}m{#1}}

\providecommand{\abs}[1]{\lvert#1\rvert}

\renewcommand{\vec}[1]{\mathbf{#1}}

\newcolumntype{L}{>{\centering\arraybackslash}m{3cm}}
\renewcommand{\vec}[1]{\mathbf{#1}}
\DeclareMathOperator*{\Assemb}{\scalerel*{\mathsf{A}}{\sum}}
\usepackage[margin=1in]{geometry}
\usepackage{lineno,hyperref}
\modulolinenumbers[5]

\journal{Underground Space}

\begin{document}

\begin{frontmatter}

\title{PHASE-FIELD MODELING OF FRACTURE FOR QUASI-BRITTLE MATERIALS}

\author{Jacinto Ulloa}
\address{a. Department of Civil Engineering, Katholieke Universiteit Leuven, B-3001 Leuven, Belgium;  b. Departamento de Recursos Hídricos y Ciencias Ambientales, Universidad de Cuenca, Cuenca 010151, Ecuador}
\cortext[JU]{Corresponding author: Jacinto Ulloa}
\ead{jacintoisrael.ulloa@kuleuven.be}
%
\author{Patricio Rodríguez}
\address{a. Universidad de Cuenca, Cuenca 010151, Ecuador; b. Institut für Kontinuumsmechanik, Leibniz Universität Hannover, Hannover 30167, Germany}
\ead{patricio.rodriguezc@ucuenca.ec}

\author{Cristóbal Samaniego}
\address{Inria Bordeaux - Sud-Ouest Research Center,
200 Avenue de la Vieille Tour, Talence 33405,
France}
\ead{cristobal.samaniego-alvarado@inria.fr}

\author{Esteban Samaniego}
\address{a. Facultad de Ingeniería, Universidad de Cuenca; b. Departamento de Recursos Hídricos y Ciencias Ambientales, Universidad de Cuenca, Cuenca 010151, Ecuador}
\ead{esteban.samaniego@ucuenca.edu.ec}

%
%
%

\begin{abstract}

This paper addresses the modeling of fracture in quasi-brittle materials using a phase-field approach to the description of crack topology. Within the computational mechanics community, several studies have treated the issue of modeling fracture using phase fields. Most of these studies have used an approach that implies the lack of a damage threshold. We herein explore an alternative model that includes a damage threshold and study how it compares with the most popular approach. The formulation is systematically explained within a rigorous variational framework. Subsequently, we present the corresponding three-dimensional finite element discretization that leads to a straightforward numerical implementation. Benchmark simulations in two dimensions and three dimensions are then presented. The results show that while an elastic stage and a damage threshold are ensured by the present model, good agreement with the results reported in the literature can be obtained, where such features are generally absent. 

\end{abstract}


\end{frontmatter}


\section{Introduction}


Crack initiation and propagation in quasi-brittle materials is both of significant scientific interest and paramount importance in engineering applications. Examples of these materials include concrete and rocks that exhibit microcracks that localize in narrow bands \citep{comi1999}. Methods developed in the framework of classical fracture mechanics, typically rooted in the pioneering works of \cite{griff}, \cite{irwin1957} and \cite{barenblatt1962}, have been relatively successfull in several engineering applications. Nevertheless, these theories present major limitations, including the inability to naturally describe crack initiation and propagation without initial defects and a prescribed crack path \citep{FrancMar1998}. Moreover, their application requires nontrivial finite element techniques such as local refinement near the crack tip \citep{nguyen2013}. In view of these drawbacks, a variety of methods have emerged in the past few decades with the objective of providing a more convenient description of fracture. 

From a continuum mechanics perspective, the study of post-critical behavior in solids can be approached using softening constitutive models. This leads to strain localization, which signals the appearance of the so-called fracture process zone. Thus, some form of representation of the high gradients in the displacement field is required to describe the material behavior after the onset of strain localization. Classical continuum mechanics cannot handle the ill-posedness of the evolution problem, thus resulting in numerical simulations with a localization zone that tends to measure zero and vanishing dissipated energy as the mesh size decreases \citep{DeBorstMull1992}. Consequently, modeling strain localization has been a constant challenge in computational mechanics \citep{OliHueSamCha2004}. To overcome these problems, the correct description of crack kinematics, topology, and energetics is required. A possible classification of these descriptions considers two subsets: i) kinematics with sharp discontinuities and ii) regularized kinematics. 


Regarding the first subset, discontinuities in the displacement field are explicitly introduced to describe sharp crack kinematics. Well-known examples of this strategy are the strong discontinuity approach (SDA) and the extended finite element method (XFEM). Building upon \cite{SimOliArm1993}, \cite{OliHueSamCha2004} proposed a continuum SDA to describe failure in geomaterials. In turn, starting with the work of \cite{MoeDolBely1999}, XFEM has been an attractive alternative for the modeling of arbitrary crack paths without remeshing. It has, for instance, been applied to describe crack growth with a cohesive law  \citep{MoeBely2002}. Moreover, a method based on local maximum-entropy (LME) interpolation was studied in conjunction with XFEM  \citep{amiri2014}. The results suggest that the method proposed therein is a competitive alternative in terms of computational cost when compared to the standard XFEM.

Although sharp crack modeling has yielded noteworthy results, major limitations can be highlighted, particularly, the inability to naturally describe complex crack topologies. In addition, the numerical treatment of crack propagation requires some form of crack-path tracking, which can be a very cumbersome task. To mitigate these difficulties, phase-field regularized models have been shown to be a competitive alternative \citep{MieWelHof2010}, which can be implemented straightforwardly. In these formulations, a continuous variable, namely, the phase-field variable, is used to describe a smooth transition between the damaged/undamaged phases.  Moreover, regularization based on phase fields can be viewed as a bridge between damage mechanics and a diffuse approximation of the sharp crack topology, which avoids the need to explicitly model discontinuities in the displacement field. In the context of fracture, the phase-field variable is represented by the scalar-valued damage quantity, whose gradient is introduced in the formulation. Thus, a clear link can be established between phase-field fracture and gradient damage. We refer to \cite{deborst2016} for a comparison between gradient-enhanced damage models and the phase-field approach to fracture.


Several studies that apply gradient-based phase-field regularizations for rate-independent systems can be  related to the work of \cite{FrancMar1998}, where a variational formulation was introduced to overcome the limitations of the Griffith model for brittle fracture, particularly, the need for a priori constraints on the crack topology. This formulation resulted in an energy functional reminiscent of the potential of \cite{MumfShah1989} for image segmentation. Subsequently, to avoid the numerical difficulties imposed by the free discontinuity problem of \cite{FrancMar1998}, \cite{BourFrancMar2000} proposed an energy functional based on phase-field regularization, where a damage gradient term was introduced (although it was not originally viewed as such). This regularization has been shown to converge to the Griffith fracture model through $\Gamma$ convergence \citep{DalToa2002}, and was inspired by the work of \cite{AmbTor1990} for the regularization of the Mumford and Shah potential. The reader is referred to \cite{BourFrancMar2008} for an overview of the regularized formulation of brittle fracture.

In the computational mechanics community, several contributions to the phase-field modeling of fracture have emerged. In \cite{MieWelHof2010}, and further developed in \cite{MieHofWel2010}, an alternative phase-field formulation was proposed, also based on thermodynamic principles. An attractive feature was incorporated in this formulation: the definition of a realistic anisotropic stored energy, obtained by defining the bulk energy density as an additive decomposition of positive (due to tension) and negative (due to compression) contributions. In this setting, damage is allowed to act on the positive component only, disallowing fracture due to compression. This formulation was extended to the dynamic case by \cite{Bord-Hugh2012}. Moreover, ductile behavior has been considered in several works. For instance, \cite{AmbGerDeL2015} combines local $J_2$ hardening plasticity with gradient damage and also considers anisotropic damage behavior. Furthermore, \cite{MieTeiAld2016,MieAldTei2016} extended these formulations to gradient plasticity combined with gradient damage. In \cite{ambati2016shells}, fracture in shells was approached using a phase-field model with isogeometric analysis \citep{hughesiso}. Phase-field modeling combined with LME interpolation was proposed by \cite{amiri2014phase} for the study of thin shells, and by \cite{amiri2016} using a fourth-order phase-field model. The above-mentioned contributions were generally built upon \cite{BourFrancMar2000}, using a particular method of approximation of the Griffith model using elliptic functionals that implies, from a mechanical perspective, the lack of a damage threshold. Therefore, an elastic stage was not included in the resulting evolution.

An interesting alternative conceived from the standpoint of continuum gradient-enhanced damage mechanics was proposed by \cite{AmorMarMau2009} and \cite{pham2011}. The general and rigorous framework of this approach is detailed in the survey of \cite{MarMauPham2016}. An important feature of this framework is that the models considered were related to the work of \citep{BourFrancMar2000}. Nonetheless, the possibility of having a damage threshold was considered, which is generally not present in most studies using phase fields to model fracture. This option was adopted by \cite{Alessi2013} and \cite{AleMarVid2014, AleMarMauVid2017}, and incorporated in a ductile fracture model. The formulation proposed therein can capture a variety of macroscopic fracture behaviors. In \cite{UllRodSam2016} and further in \cite{rodriguez2018}, within the same framework, hardening effects and gradient plasticity with variable internal length were considered. The variational approach used in these studies follows a rigorous energetic formulation, formalized by  \cite{Mielke2006} and \cite{Mielke2015}. As discussed in \cite{Alessi2016}, the energetic formulation presents several advantages with respect to classical formulations. For instance, the governing equations are naturally derived using the calculus of variations and three physical principles: the stability condition, energy balance and the fulfilment of the second law of thermodynamics. Moreover, the definition of a global energy functional leads to a robust numerical implementation that can be solved using a simple staggered scheme.

In this study, we perform an analysis of this alternative method to describe fracture using phase fields for quasi-brittle materials. In Section 2, the governing equations of the problem are systematically derived following the variational framework. Then, Section 3 presents the details of the corresponding finite element implementation. The numerical simulations are presented in Section 4 and are compared with the results reported in the literature, where the most popular model for quasi-brittle materials was used.

\section{Formulation}

We adopt an energetic framework for the description of the behavior of deformable solids in the rate-independent case \citep{Mielke2006}. In this work, we assume evolutions under small strains, with the exception of certain localized regions. Our goal is to describe quasi-brittle fracture with an elastic stage, which results in a two-field formulation, considering displacements $\boldsymbol{u}$ and the damage variable $\alpha$ as primary fields. 

Following the theory of generalized standard materials \citep{halphen1975}, an energy functional is defined as the sum of potential and dissipative energy terms:

\begin{equation}
\mathcal{W}(\boldsymbol{u},\alpha)=\mathcal{P}(\boldsymbol{u},\alpha)+\mathcal{D}(\alpha).
\label{energyfunct0}
\end{equation}
\\
The minimization of this functional with respect to $\boldsymbol{u}$ and $\alpha$ separately entails the fulfillment of the momentum balance and damage criterion, respectively. The weak form of each of these equations is naturally obtained and can be discretized using the finite element method. It is noteworthy that the definition of a total energy quantity is not always straightforward: it strongly depends on the form of the dissipated work $\mathcal{D}(\alpha)$  \citep{AleMarMauVid2017}. 

The fact that gradient-enhanced damage models entail a regularization of the softening problem that leads to mesh-independent solutions with nonvanishing dissipation is well known. Our approach follows the process of phase-field models, which are strongly linked to gradient-enhanced damage. In fact, the primary difference between both formulations is how they are conceived: the primary idea of phase-field models is the description of the discontinuity of the crack using a continuous field, while  gradient damage models are approached from a mechanical perspective, where gradients are included to regularize the ill-posed boundary value problem. Nevertheless, an important difference is that models developed in a purely gradient-damage framework typically result in broadening of the damage zone, whereas phase-field models capture a sharp transition zone in more naturally owing to the use of a degradation function \citep{deborst2016}. Thus, in this study, the phase-field approach is employed, but the constitutive functions are defined from the perspective of damage mechanics to preserve the mechanical interpretation. 

The ingredients to establish the energetic formulation are described in the following sections.

\subsection{Primary and state variables}\

\begin{figure}
\centering
\includegraphics[scale=1.1]{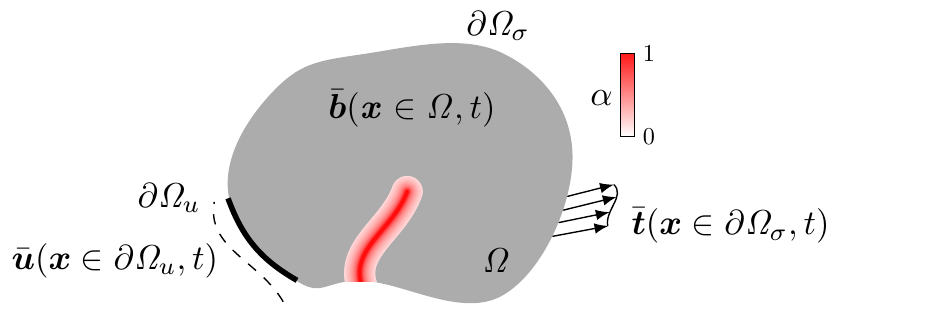}
\caption{\ Diffuse crack description.}
\label{solid}
\end{figure}  

Consider a solid body ${\mathit{\Omega}}$ with Neumann boundary $\partial{\mathit{\Omega}}_\sigma$ and Dirichlet boundary $\partial{\mathit{\Omega}}_u$. Let $\boldsymbol{u}(\boldsymbol{x},t)$ be the displacement of a point $\boldsymbol{x} \in {\mathit{\Omega}}$ at time $t$. Because the small strain hypothesis is adopted in this work, no distinction is made between the original and the deformed configuration of the body. Consequently, the second-order total strain tensor is defined as 

\begin{equation}
\boldsymbol{\epsilon}=\boldsymbol{\epsilon}(\boldsymbol{u})=\frac{1}{2}\bigg(\nabla\boldsymbol{u}+\nabla\boldsymbol{u}^{\mathrm{T}}\bigg),
\end{equation}
\\
where we have dropped the explicit dependence of the involved fields on $\boldsymbol{x}$ and $t$ for the sake of notational simplicity.

For the phase-field description of the crack topology, the internal scalar-valued damage variable $\alpha$ is characterized by

\begin{equation}
\alpha \in [0,1] \ \  \text{with} \ \  \dot{\alpha}\geq0,
\end{equation}
\\
which indicates the damaged/undamaged points in the solid. A value of $\alpha = 0$ corresponds to an undamaged material state, while $\alpha=1$ defines a completely broken material state. In the formulation presented herein, regularization is attained using the gradient of the damage variable $\nabla\alpha$, which allows for the formation of a diffuse crack whose width is finite and depends on the damage internal length scale $\eta$. Figure~\ref{solid} shows this type of crack description within the general problem setting.

The above-mentioned variables are fundamental in the multifield model. As summarized in Table~\ref{Table1}, the primary global variables for the model are $\boldsymbol{u}$ and $\alpha$. Meanwhile, the constitutive state variables, which define the material behavior of each point within the solid, are $\boldsymbol{\epsilon}$, $\alpha$ and $\nabla\alpha$ (see Table~\ref{Table2}).

\begin{table}[]
\centering
\caption{\ Global primary variables}
{\begin{tabular}{llll}
\multicolumn{2}{c}{Primary Variable}                                                              & Field                          & Type                                        \\ \hline
$\boldsymbol{u}$               & displacement field                                   & vector                         & observable                                 \\
$\alpha$                       & damage                                               & scalar                         & internal                                   \\ 
\hline
\end{tabular}}
\label{Table1}
\end{table}

\begin{table*}[]
\centering
\caption{\ Constitutive state variables}
{\begin{tabular}{llll}
\multicolumn{2}{c}{State Variable}                                                              & Field                          & Type                                        \\ \hline
$\boldsymbol{\epsilon}$               & total strain                                    & second-order tensor                         & observable                                  \\
$\alpha$                       & damage                                               & scalar                         & internal                                    \\ 
$\nabla\alpha$                 & damage gradient                                      & vector                         & internal                                   
 \\  
 \hline
\end{tabular}}
\label{Table2}
\end{table*}

\subsection{Variational approach}\

\subsubsection{Energy functional}\

This section describes the energetic formulation in the context of gradient damage. Adopting the theory of generalized standard materials \citep{halphen1975, germ}, the variational approach to fracture begins with the description of the basic energetic quantities. The global stored energy is defined using the elastic energy density $\Psi$ as

\begin{equation}
\label{stored_elast}
\mathcal{E}(\boldsymbol{\epsilon},\alpha)=\int_{{\mathit{\Omega}}}\Psi(\boldsymbol{\epsilon,\alpha})\mathrm{d}{\mathit{\Omega}}.
\end{equation}
\\
The Cauchy stress tensor $\boldsymbol{\sigma}$ is given by the constitutive equation

\begin{equation}
\boldsymbol{\sigma}(\boldsymbol{\epsilon},\alpha)=\frac{\partial\Psi(\boldsymbol{\epsilon,\alpha})}{\partial\boldsymbol{\epsilon}}.
\end{equation}
\\
The dissipative nature of the internal variables is characterized by the definition of the dissipation potential \citep{AleMarMauVid2017}. Considering rate-independence, the  dissipation potential is a first-order homogeneous convex function of the rate of the damage variable, and can be expressed as

\begin{equation}
\label{totaldisspot}
\begin{aligned}
&\Phi = w'(\alpha)\dot{\alpha}+\partial_{t}\left(\frac{1}{2}\eta^2\nabla\alpha\cdot\nabla\alpha\right).\\
\end{aligned}
\end{equation}
\\
The function $w(\alpha)$ represents the dissipated energy in a homogeneous damage process. In addition, the term corresponding to the damage gradient with its corresponding internal length scale is included. This parameter is directly related to the size of the localization zone, this implying several advantages, including the ability to describe structural stability and size effects \citep{MarMauPham2016}. With Eq.~\eqref{totaldisspot}, assuming a smooth evolution, the dissipated work $\mathcal{D}$ follows:

\begin{equation}
\label{dis_work}
\mathcal{D}(\alpha)=\int_{\mathit{\Omega}}\left(w(\alpha)+\frac{1}{2}\eta^2\nabla\alpha\cdot\nabla\alpha\right)\mathrm{d}{\mathit{\Omega}}.
\end{equation}
\\
In this work, no external forces are considered (although their inclusion would be straightforward), and displacements are imposed on $\partial{\mathit{\Omega}}_u$. Thus, the total energy functional is defined using Eq.~\eqref{dis_work} and Eq.~\eqref{stored_elast} as

\begin{equation}
\label{energyfunct}
\mathcal{W}(\boldsymbol{u},\alpha)=\mathcal{E}(\boldsymbol{u},\alpha)+\mathcal{D}(\alpha)=\int_{{\mathit{\Omega}}}\Psi(\boldsymbol{\epsilon},\alpha)\mathrm{d}{\mathit{\Omega}}+\int_{\mathit{\Omega}}\left(w(\alpha)+\frac{1}{2}\eta^2\nabla\alpha\cdot\nabla\alpha\right)\mathrm{d}{\mathit{\Omega}}.
\end{equation}
\\
\subsubsection{Energetic formulation}\

The building blocks of the variational framework adopted in this work are the following principles \citep{pham2011,Alessi2013}:

\begin{enumerate}
\item Stability condition,
\item Energy balance,
\item Irreversibility condition.
\end{enumerate}

The irreversibility condition is imposed on the damage variable to disallow material healing. It is applied numerically by considering the damage value corresponding to the previous load step as the minimum admissible level of damage for a given position in the body as follows:

\begin{equation}
\alpha_{t_{n+1}}(\boldsymbol{x})=
\begin{dcases}
    \alpha_{{n+1}}(\boldsymbol{x}),& \text{if} \ \ \alpha_{{n+1}}(\boldsymbol{x})\geq\alpha_{{n}}(\boldsymbol{x})\\
    \alpha_{{n}}(\boldsymbol{x}),& \text{if}  \ \ \alpha_{{n+1}}(\boldsymbol{x})<\alpha_{{n}}(\boldsymbol{x}),\\
\end{dcases}
\label{condirralg}
\end{equation}
\\
where $\alpha_{{n+1}}(\boldsymbol{x})$ is the damage value for the current time step ${n+1}$ in  $\boldsymbol{x} \in {\mathit{\Omega}}$. 
\vspace{0.4cm}
\paragraph{\bfseries{Stability condition}}\
A directional stability condition may be defined by the Taylor expansion of the (sufficiently regular) energy functional. The first-order variations of Eq.~\eqref{energyfunct} yield the first-order stability condition:

\begin{equation}
\left.\frac{\mathrm{d}}{\mathrm{d}h}\mathcal{W}(\boldsymbol{u}+h\tilde{\boldsymbol{u}},\alpha+h\tilde{\alpha})\right|_{h=0}=\int_{\mathit{\Omega}}\bigg[\boldsymbol{\sigma}(\boldsymbol{\epsilon},\alpha):{\boldsymbol{\epsilon}}(\tilde{\boldsymbol{u}})+\bigg(\frac{\partial\Psi(\boldsymbol{\epsilon,\alpha})}{\partial{\alpha}}+w'(\alpha)\bigg)\tilde{\alpha}+\eta^2\nabla\alpha\cdot\nabla\tilde{\alpha}\bigg]\mathrm{d}{\mathit{\Omega}}\geq0.
\label{stab}
\end{equation}
\\
We refer the reader to \cite{pham_mara}, \cite{pham2011}, and  \cite{MarMauPham2016} for more details on the stability condition in the context of gradient-enhanced damage and to \cite{Mielke2006} and \cite{Mielke2015} for more general definitions.

From Eq.~\eqref{stab}, the following results are obtained.

\begin{itemize}
\item For $\tilde{\alpha}=0$:

\begin{equation}
\label{weak_u}
\int_{\mathit{\Omega}}\boldsymbol{\sigma}(\boldsymbol{\epsilon},\alpha):{\boldsymbol{\epsilon}}(\tilde{\boldsymbol{u}})\mathrm{d}{\mathit{\Omega}}=0,
\end{equation}
\\
representing the weak form of the equilibrium equation in the absence of external loads.
\\
\item For $\tilde{\boldsymbol{u}}=\boldsymbol{0}$:

\begin{equation}
\label{weak_a}
\int_{\mathit{\Omega}}\bigg[\left(\frac{\partial\Psi(\boldsymbol{\epsilon,\alpha})}{\partial{\alpha}}+w'(\alpha)\right)\tilde{\alpha}+\eta^2\nabla\alpha\cdot\nabla\tilde{\alpha}\bigg] \mathrm{d}{\mathit{\Omega}}\geq0,
\end{equation}
\\
in which the weak form of the damage criterion is obtained. The last term is integrated by parts and the gradient-dependent damage yield criterion is recovered in the local form

\begin{equation}
\label{yield_a}
\begin{aligned}
f_{d}(\boldsymbol{u},\alpha)=&-\frac{\partial\Psi(\boldsymbol{\epsilon,\alpha})}{\partial{\alpha}}-w'(\alpha)+\eta^2\nabla\cdot\nabla\alpha\leq0 \quad \text{in} \quad {\mathit{\Omega}}.
\end{aligned}
\end{equation}
\\
Moreover, it can be easily shown that the boundary conditions 

\begin{equation}
\boldsymbol{n}\cdot\nabla\alpha \geq 0 \quad \text{on} \quad \partial{\mathit{\Omega}}
\label{bc1}
\end{equation}
\\
also follow, where $\boldsymbol{n}$ is the unit outward normal vector to $\partial{\mathit{\Omega}}$. 

\end{itemize}
\vspace{0.4cm}
\paragraph{\bfseries{Energy balance}}\
The energy balance states the need for the total energy to remain constant as the state variables evolve. Thus, it is essentially a restatement of the first law of thermodynamics. Following a procedure analogous to the treatment of the stability condition, the energy balance leads to 

\begin{equation}
\begin{aligned}
&\int_{\mathit{\Omega}}\bigg[\boldsymbol{\sigma}(\boldsymbol{\epsilon},\alpha):{\boldsymbol{\epsilon}}(\dot{\boldsymbol{u}})+\bigg(\frac{\partial\Psi(\boldsymbol{\epsilon,\alpha})}{\partial{\alpha}}+w'(\alpha)-\eta^2\nabla\cdot\nabla\alpha\bigg)\dot{\alpha}\bigg]\mathrm{d}{\mathit{\Omega}}=0.
\end{aligned}
\end{equation}
\\
For $\dot{\boldsymbol{u}}=\boldsymbol{0}$, and using Eq.~\eqref{yield_a}, the damage consistency conditions

\begin{equation}
f_{d}(\boldsymbol{u},\alpha)\dot{\alpha}=0 \quad \text{in} \quad {\mathit{\Omega}}
\label{const_a}
\end{equation}
\\
are obtained, along with the boundary conditions

\begin{equation}
\dot{\alpha}\boldsymbol{n}\cdot\nabla\alpha = 0 \quad \text{on} \quad \partial{\mathit{\Omega}}.
\end{equation}
\\
Thus, for any $\dot{\alpha} > 0$,

\begin{equation}
\boldsymbol{n}\cdot\nabla\alpha = 0 \quad \text{on} \quad \partial{\mathit{\Omega}},
\end{equation}
\\
representing the additional boundary conditions required for the gradient-enhanced evolution equations. 

Eqs. \eqref{yield_a}, \eqref{const_a} and the irreversibility condition represent gradient-enhanced versions of the evolution equations of classical local models in the form of Karush-Kuhn-Tucker conditions.

\subsection{Constitutive assumptions}\
\label{constitutive}

This section defines the constitutive assumptions for the damaging process. The following definitions represent the functions and parameters that are intended to capture the behavior of quasi-brittle materials. The resulting model should describe the softening behavior and strain localization as the phase-field variable evolves. 

The initial elastic energy density, corresponding to a sound material state, is expressed as

\begin{equation}
\begin{aligned}
\label{ElastEnerDens}
\Psi_0(\boldsymbol{\epsilon})&=\frac{1}{2}\boldsymbol{\epsilon}:\boldsymbol{\sigma}_0=\frac{1}{2}\boldsymbol{\epsilon}:\mathbf{C}_0:\boldsymbol{\epsilon}=\frac{1}{2}\lambda\tr(\boldsymbol{\epsilon})^2+\mu(\boldsymbol{\epsilon}:\boldsymbol{\epsilon}),
\end{aligned}
\end{equation}
\\
where $\boldsymbol{\sigma}_0$ is the undamaged Cauchy stress tensor. The symbols $\lambda$ and $\mu$ are the Lamé parameters, and $\mathbf{C}_0$ is the undamaged elastic fourth-order tensor, given for an isotropic elastic material by

\begin{equation}
\mathbf{C}_0=\lambda\boldsymbol{1}\otimes\boldsymbol{1}+2\mu\boldsymbol{I},
\end{equation}
\\
where $\boldsymbol{1}$ is the second-order identity tensor and $\boldsymbol{I}$ is the fourth-order symmetric identity tensor.

A differentiated damaging behavior is considered by decomposing the elastic energy density into positive (due to tension) and negative (due to compression) energies. Damage is allowed to act on the positive part of the elastic energy only, thus disallowing failure due to compression. Consequently, fracture only takes place in regions under tension, while crack interpenetration is disallowed in the compressed regions \citep{AmorMarMau2009}. This energy split allows to describe quasi-brittle fracture in materials that exhibit different strengths in tension and compression. Examples of these are concrete and other geomaterials, where cracking is associated with tensile stresses. A decomposition of stress into tension and compression was carried out by \cite{faria2004} to develop a local damage model for concrete structures under cyclic loading conditions. In the context of phase-field fracture, \cite{AmorMarMau2009} proposed an energy decompositon based on a volumetric-deviatoric split. Alternative approaches have been proposed in the literature, such as a spectral decomposition by \cite{MieWelHof2010}, a no-tension split by \cite{freddi2010}, and recently, a stress-based split by \cite{steinke2018}. Although each approach owns advantages in certain cases, the definition of an optimal split remains an open issue. In this work, we adopt the volumetric-deviatoric model of \cite{AmorMarMau2009}. Therein, the elastic energy density is expressed in terms of the volumetric and deviatoric components, and the positive and negative contributions are expressed as  

\begin{equation}
\label{TensCompDecomp}
\Psi^{+}_0(\boldsymbol{\epsilon})=\frac{1}{2}K\langle\tr(\boldsymbol{\epsilon})\rangle_+^2+\mu(\boldsymbol{\epsilon}_{\mathrm{dev}}:\boldsymbol{\epsilon}_{\mathrm{dev}}) \quad \text{and} \quad
\Psi^{-}_0(\boldsymbol{\epsilon})=\frac{1}{2}K\langle\tr(\boldsymbol{\epsilon})\rangle_-^2,
\end{equation}
\\
where the ramp function $\langle\tr(\boldsymbol{\epsilon})\rangle_\pm=\frac{1}{2}\big(\tr(\boldsymbol{\epsilon})\pm\abs{\tr(\boldsymbol{\epsilon})}\big)$ is used, $K=\lambda+\frac{2}{3}\mu$ is the bulk modulus and $\boldsymbol{\epsilon}_{\mathrm{dev}}$ denotes the deviatoric part of the strain tensor. Anisotropic material degradation can now be described by a stored energy density of the form 

\begin{equation}
\Psi(\boldsymbol{\epsilon},\alpha)=f(\alpha)\Psi_0^{+}(\boldsymbol{\epsilon})+\Psi_0^{-}(\boldsymbol{\epsilon}).
\end{equation}
\\
Using the Heaviside step function $\heavi$, the Cauchy stress tensor can be decomposed as follows:

\begin{equation}
\label{StressDecomp}
\begin{aligned}
\boldsymbol{\sigma}(\boldsymbol{\epsilon},\alpha)=\frac{\partial\Psi}{\partial\boldsymbol{\epsilon}}=&\underbrace{\bigg(K\heavi\big(\tr(\boldsymbol{\epsilon})\big)\tr(\boldsymbol{\epsilon})\boldsymbol{1}+2\mu\boldsymbol{\epsilon}_{\mathrm{dev}}\bigg)}_{ \textstyle\boldsymbol{\sigma}_0^+}f(\alpha)+\underbrace{K\heavi\big(-\tr(\boldsymbol{\epsilon})\big)\tr(\boldsymbol{\epsilon})\boldsymbol{1}}_{ \textstyle\boldsymbol{\sigma}_0^-}.
\end{aligned}
\end{equation}
\\
Regarding the elastic energy degradation, we require that 

\begin{equation}
 f'(\alpha) < 0 \quad \text{and} \quad f(1) = 0.
\end{equation}
\\
The first condition ensures material degradation, while the second defines a fractured material state. For this purpose, we adopt the quadratic function of damage that has been widely used in phase-field models in the literature:

\begin{equation}
f(\alpha) = (1-\alpha)^2.
\end{equation}
\\
The main advantage of using a quadratic degradation function is that the damage criterion is a (constrained) linear partial differential equation, as will be clear in the following section. Advantages of using a cubic degradation function were reported by \cite{Bord-Hugh2016}; however, we adopt the quadratic function to maintain our claim of a simple numerical implementation. 

The dissipation due to local damage evolution is defined by the positive-valued function $w(\alpha)$ that represents the dissipated energy of a volume element throughout the damage process. As described by \cite{MarMauPham2016}, this dissipation is set to increase to a critical value 

\begin{equation}
w(1)=w_0<\infty.
\end{equation}
\\
Among the existing phase-field models, two alternatives can be found for the local damage dissipation:

\begin{equation}
\label{DamModels}
w(\alpha) =
\begin{dcases}
     w_0\alpha & \text{model with an elastic stage},\\
     w_0\alpha^2 & \text{model without an elastic stage.}\\
\end{dcases}
\end{equation}
\\
The critical damage dissipation $w_0>0$ represents the energy dissipated during a complete damage process for a volume element, and is related to the fracture toughness $G_\mathrm{c}$ used in other formulations by \citep{MarMauPham2016}:

\begin{equation}
\label{toughness}
G_\mathrm{c} = 2l\int_0^1\sqrt{2w_0w(\beta)}\mathrm{d}\beta=c_\mathrm{w}\frac{lw_0}{\sqrt{2}} \ \ \ \text{with} \ \  c_\mathrm{w} =4\int_0^1\sqrt{\frac{w_0}{2w(\beta)}}\mathrm{d}\beta,
\end{equation}
\\
where the internal length is denoted by  $l$ and can be expressed in terms of $\eta$ as
\begin{equation}
l = \frac{\eta}{\sqrt{w_0}}.
\end{equation}
\\
From the previous expressions and the energetic definitions \eqref{dis_work} and \eqref{energyfunct}, note that $\eta$ has the units of $\text{energy}^{1/2}\text{length}$.  

The model without an elastic stage has been consistently used in phase-field models, as shown in \cite{AmbGerDeL2015, AmbKruDel2016} and \cite{Bord-Hugh2012,Bord-Hugh2016}. For this choice, the absence of an elastic stage is due to $w'(0)=0$. In this work, we adopt $w(\alpha)=w_0\alpha$ as the constitutive choice, for which $w'(0)=w_0>0$. Therefore, this function allows for the existence of an initial elastic stage before the damage evolution. Moreover, the parameter $w_0$ can be considered as a damage threshold from the viewpoint of classical continuum damage mechanics. We consider this an attractive feature because it forbids the onset of damage before a critical stress state is reached. From this choice and using Eq.~\eqref{toughness}, $w_0$ can be related to the fracture toughness by

\begin{equation}
G_\mathrm{c} = \frac{4\sqrt{2}}{3}w_0l.
\end{equation} 

With these definitions, the damage criterion from Eq.~\eqref{yield_a} becomes 

\begin{equation}
\label{yield_a2}
f_{\mathrm{d}}(\boldsymbol{u},\alpha)=(1-\alpha)\boldsymbol{\sigma}_0^+:(\boldsymbol{\epsilon})-w_0+\eta^2\nabla\cdot\nabla\alpha\leq0.
\end{equation}

%
%
%

\label{ConstAssmp}

\subsection{Alternate minimization}\

With the definition of the main ingredients of the variational approach, a numerical solution can be readily obtained. For this purpose, an alternate minimization algorithm is applied, which naturally emerges from the energetic principles. This procedure takes advantage of the fact that although the global energy is nonconvex, it is convex with respect to $\boldsymbol{u}$ and $\alpha$ individually. Introducing the constitutive assumptions of the previous section into Eq.~\eqref{energyfunct}, the global energy functional reads 

\begin{equation}
\label{energyfunct2}
\mathcal{W}(\boldsymbol{u},\alpha)=\int_{{\mathit{\Omega}}}\frac{1}{2}(1-\alpha)^2\boldsymbol{\epsilon}:\boldsymbol{\sigma}_0^+\mathrm{d}{\mathit{\Omega}}+\int_{{\mathit{\Omega}}}\left(w_{0}\alpha+\frac{1}{2}\eta^2\nabla\alpha\cdot\nabla\alpha\right)\mathrm{d}{\mathit{\Omega}}.
\end{equation}
\\
The alternate minimization follows.

\begin{itemize}
\item Minimization with respect to the displacement field:

\begin{equation}
\label{displeq}
\left.\frac{\mathrm{d}}{\mathrm{d}h}\mathcal{W}(\boldsymbol{u}+h\tilde{\boldsymbol{u}},\alpha)\right|_{h=0} =\int_{\mathit{\Omega}}\big((1-\alpha)^2\boldsymbol{\sigma}_0^++\boldsymbol{\sigma}_0^-\big):{\boldsymbol{\epsilon}}(\tilde{\boldsymbol{u}})\mathrm{d}{\mathit{\Omega}}=0. 
\end{equation}
\\
\item Minimization with respect to the damage field:

\begin{equation}
\label{dameq}
\left.\frac{\mathrm{d}}{\mathrm{d}h}\mathcal{W}(\boldsymbol{u},\alpha+h\tilde{\alpha})\right|_{h=0}=\int_{\mathit{\Omega}}\bigg((1-\alpha)\boldsymbol{\epsilon}:\boldsymbol{\sigma}_0^+\tilde{\alpha}-w_0\tilde{\alpha}-\eta^2\nabla\alpha\cdot\nabla\tilde{\alpha}\bigg)\mathrm{d}{\mathit{\Omega}}=0.
\end{equation}
\end{itemize}

The last two expressions can be easily viewed as weak forms of the Euler equations of the underlying global minimization problem. In this work, the staggered solution is applied incrementally, which results in a straightforward implementation, given the choice of constitutive functions. The solution of the problem is subsequently completed with the imposition of the irreversibility condition in the algorithmic form \eqref{condirralg}. The numerical solution is described in the following section. 

\section{Numerical solution and implementation}
This section presents a straightforward implementation of the numerical solution of Eqs.~\eqref{displeq} and \eqref{dameq}. Linear tetrahedral finite elements are used throughout this section, although the use of higher-order elements would be equally straightforward. 

The overall procedure can be summarized as follows: Eqs.~\eqref{displeq} and \eqref{dameq} are first written in discrete form using the Voigt notation by projection over finite elements. The elastic equilibrium equation is subsequently used to obtain the displacement field, and the damage field is obtained using the updated displacements. This process is repeated iteratively to approximate the two primary fields at a specific pseudo-time. The discrete versions are then solved at the following pseudo-time, entering a temporal incremental scheme.    

\subsection{Numerical solution}

\subsubsection{Finite element approximation}\

The displacement field approximation is defined as

\begin{equation}
\boldsymbol{u}^\mathrm{h}=\vec{N}_\mathrm{v}\vec{u},
\end{equation}
\\
where the global vector of nodal displacements $\vec{u}$ and the shape function matrix $\vec{N}_{\mathrm{v}}$ are obtained using an assembly operator $\Assemb$:

\begin{equation}
\vec{u}= \Assemb_{{el}=1}^n \vec{u}^{el}, \quad \vec{u}^{el}=\begin{bmatrix} \vec{u}_1^{el} & \vec{u}_2^{el} & \vec{u}_3^{el} & \vec{u}_4^{el} \end{bmatrix}^\mathrm{T},  
\end{equation}
\\
and

\begin{equation}
\vec{N}_\mathrm{v}=\Assemb_{{el}=1}^n {\begin{bmatrix} \vec{N}_\mathrm{v}^{1} & \vec{N}_\mathrm{v}^{2} & \vec{N}_\mathrm{v}^{3} & \vec{N}_\mathrm{v}^{4} \end{bmatrix}}^{el}, \quad \vec{N}_\mathrm{v}^i= \begin{bmatrix} N_i & 0 & 0 \\ 0 & N_i & 0 \\ 0 & 0 & N_i \end{bmatrix}. 
\end{equation}
\\
The superscript ${el}$ denotes an element vector or matrix, and the index $i\in[1,2,3,4]$ indicates the node number with $N_i$ representing the corresponding shape functions. The vector $\vec{u}_i^{el}$ contains the nodal displacements in each spatial direction. 

The strain vector inside an element is defined as

\begin{equation}
\label{strainapp}
\boldsymbol{\upvarepsilon}^{el}=\begin{bmatrix}
\epsilon_x^{el} & \epsilon_y^{el} & \epsilon_z^{el} & \gamma_{xy}^{el} & \gamma_{yz}^{el} & \gamma_{zx}^{el}\end{bmatrix}^\mathrm{T},\\
\end{equation}
\\
where $\gamma$ represents the shear strain components. The global strain vector is obtained from the nodal displacements as

\begin{equation}
\label{strainapp1}
\boldsymbol{\upvarepsilon}=\vec{B}_\mathrm{v}\vec{u},\\
\end{equation}
\\
where

\begin{equation}
\vec{B}_\mathrm{v}=\Assemb_{{el}=1}^n {\begin{bmatrix} \vec{B}_\mathrm{v}^1 & \vec{B}_\mathrm{v}^2 & \vec{B}_\mathrm{v}^3 & \vec{B}_\mathrm{v}^4 \end{bmatrix}}^{el}, \ \ \ \ \vec{B}_\mathrm{v}^i= \begin{bmatrix} \partial N_i/\partial x & 0 & 0 & \partial N_i/\partial y & 0 & \partial N_i/\partial z \\ 0 & \partial N_i/\partial y & 0 & \partial N_i/\partial x & \partial N_i/\partial z & 0 \\0 & 0 & \partial N_i/\partial z & 0 & \partial N_i/\partial y & \partial N_i/\partial x \end{bmatrix} ^\mathrm{T}.
\end{equation}
\\
The damage and gradient damage fields are approximated by

\begin{equation}
\label{scalarapp}
{\alpha}^\mathrm{h}=\vec{N}_\mathrm{s}\boldsymbol{\alpha}
\end{equation}
\\
and

\begin{equation}
\label{scalargradapp}
\nabla\alpha^\mathrm{h} = \vec{B}_\mathrm{s}\boldsymbol{\alpha},
\end{equation}
\\
where the assembly operator is used again:

\begin{equation}
\boldsymbol{\alpha}= \Assemb_{{el}=1}^n \boldsymbol{\alpha}^{el}, \ \ \ \ \boldsymbol{\alpha}^{el}=\begin{bmatrix} \alpha_1^{el} & \alpha_2^{el} & \alpha_3^{el} &\alpha_4^{el} \end{bmatrix} ^\mathrm{T},
\end{equation}
\\
and

\begin{equation}
\vec{N}_\mathrm{s}=\Assemb_{{el}=1}^n {\begin{bmatrix} N_1 & N_2 & N_3 & N_4 \end{bmatrix}}^{el}, \ \ \ \ \vec{B}_\mathrm{s}=  \Assemb_{{el}=1}^n {\begin{bmatrix} \vec{B}_\mathrm{s}^1 & \vec{B}_\mathrm{s}^2 & \vec{B}_\mathrm{s}^3 & \vec{B}_\mathrm{s}^4 \end{bmatrix}}^{el}, \ \ \ \ \vec{B}_\mathrm{s}^i= \begin{bmatrix} \partial N_i / \partial x & \partial N_i / \partial y & \partial N_i / \partial z \end{bmatrix}^\mathrm{T}.
\end{equation}
\\
Hereinafter, all element vectors and matrices are assumed to be assembled into their corresponding global forms through $\Assemb$. 

%

\subsubsection{Discrete forms}\

The discrete versions of the evolution equations consist of systems of linear equations, from where the solution to each primary variable can be easily obtained. Before presenting the discrete forms, we introduce the definitions that follow. The volumetric strain is defined as 

\begin{equation}
\epsilon_\mathrm{v}=\frac{1}{3}(\epsilon^x+\epsilon^y+\epsilon^z),
\end{equation}
\\
and the deviatoric part of the strain vector inside an element reads 

\begin{equation}
\label{devstrain}
\boldsymbol{\upvarepsilon}_{\mathrm{dev}}^{el}=\begin{bmatrix}
	\epsilon_x^{el}-\epsilon_\mathrm{v}^{el} & \epsilon_y^{el}-\epsilon_\mathrm{v}^{el} & \epsilon_z^{el}-\epsilon_\mathrm{v}^{el} & \dfrac{\gamma_{xy}^{el}}{2} & \dfrac{\gamma_{yz}^{el}}{2} & \dfrac{\gamma_{zx}^{el}}{2}
	\end{bmatrix}^\mathrm{T}.
\end{equation}
\\
For the anisotropic model, the constitutive matrix is expressed in terms of the bulk and shear moduli. Introducing the volumetric and deviatoric operators

\begin{equation}
\mathbb{P}_\mathrm{V}=\begin{bmatrix}
1 & 1 & 1 & 0&0& 0 \\
1 & 1 & 1 & 0&0& 0 \\
1 & 1 & 1 &  0&0& 0 \\
0 & 0 & 0 & 0&0& 0 \\
0 & 0 & 0 & 0&0& 0 \\
0 & 0 & 0 & 0&0& 0 \\
\end{bmatrix}  \quad \text{and} \quad \mathbb{P}_\mathrm{D}=\begin{bmatrix}
\frac{2}{3} & -\frac{1}{3} & -\frac{1}{3} & 0 & 0 & 0\\
-\frac{1}{3} & \frac{2}{3} & -\frac{1}{3} & 0 & 0 & 0 \\
-\frac{1}{3} & -\frac{1}{3} & \frac{2}{3} &  0 & 0 & 0 \\
0 & 0 & 0 & \frac{1}{2} & 0 & 0 \\
0 & 0 & 0 & 0 & \frac{1}{2} & 0 \\
0 & 0 & 0 & 0 & 0 & \frac{1}{2}
\end{bmatrix},
\end{equation}
\\
the local elastic matrix is defined in the decomposed form

\begin{equation}
\label{constmat}
\mathbf{D}^{el} = K\big[\heavi(\epsilon_\mathrm{v}^{el})(1-\vec{N}_\mathrm{s}^{el}\boldsymbol{\alpha}^{el})^2+\heavi(-\epsilon_\mathrm{v}^{el})\big]\mathbb{P}_\mathrm{V}+2\mu(1-\vec{N}_\mathrm{s}^{el}\boldsymbol{\alpha}^{el})^2\mathbb{P}_\mathrm{D},
\end{equation}
\\
where $\heavi$ is the Heaviside step function. The global damaged stress vector can than be readily obtained from

\begin{equation}
\label{vectstress}
\boldsymbol{\upsigma}=\mathbf{D}\boldsymbol{\upvarepsilon}.
\end{equation}

Using the previous definitions, the discrete form of Eq.~\eqref{displeq} can be expressed globally as

\begin{equation}
\int_{{\mathit{\Omega}}}{\vec{B}_\mathrm{v}}^\mathrm{T}\mathbf{D}\vec{B}_\mathrm{v}\mathrm{d}{\mathit{\Omega}}\vec{u}=\vec{0},
\label{sol_u}
\end{equation}
\\
from where the global displacement vector $\vec{u}$ can be readily obtained after imposing boundary conditions.

Finally, the global damage field vector $\boldsymbol{\alpha}$ is obtained from the discrete version of Eq.~\eqref{dameq}, which is given by

\begin{equation}
\int_{\mathit{\Omega}}\bigg[{\vec{N}_\mathrm{s}}^\mathrm{T}\bigg((\vec{B}_\mathrm{v}\vec{u})^\mathrm{T}\boldsymbol{\upsigma}_0^+\bigg)\vec{N}_\mathrm{s}+{\vec{B}_\mathrm{s}}^\mathrm{T}\eta^2\vec{B}_\mathrm{s}\bigg]\boldsymbol{\alpha}=\\
\int_{\mathit{\Omega}}\vec{N}_\mathrm{s}^\mathrm{T}\bigg((\vec{B}_\mathrm{v}\vec{u})^\mathrm{T}\boldsymbol{\upsigma}_0^+-w_0\bigg)\mathrm{d}{\mathit{\Omega}},
\label{sol_a}
\end{equation}
\\
where the element form of $\boldsymbol{\upsigma}_0^+$ reads

\begin{equation}
{\boldsymbol{\upsigma}_0^{+}}^{el}=K\heavi(\epsilon_\mathrm{v}^{el})\tr(\boldsymbol{\upvarepsilon}^{el})\begin{bmatrix}
1 & 1 & 1 & 0 & 0 & 0
\end{bmatrix}^\mathrm{T} + 2\mu\boldsymbol{\upvarepsilon}_{\mathrm{dev}}^{el}.
\end{equation}
\\

\subsection{Implementation details}\

The implementation of the proposed model is summarized in this section. We emphasize that our goal in this study is not improving computational efficiency, but to present a clear and straightforward implementation (in the spirit of \cite{AlbCarSte1999}) of a rather complex problem.

The overall architecture of the proposed implementation consists of the following modules:

\begin{itemize}
\item Preprocessing: generating the mesh. Details on mesh generation are outside the scope of this paper.  
\item Data input: reading data files and defining material constants, convergence factors and imposed displacements.
\item Initialization: initializing convergence vectors and  storage matrices.
\item One module for each primary variable: obtaining the global vectors $[\vec{u}, \boldsymbol{\alpha}]$ using two separate finite-element problems. 
\item Main module: it sequentially calls the modules for each primary variable into an iterative procedure for each load step. 
\item Postprocessing: displaying the primary fields illustratively. 
\end{itemize}

\subsubsection{Modules for primary variables}\

The nodal vectors corresponding to the global primary fields, $\vec{u}$ and $\boldsymbol{\alpha}$, are obtained in separate modules: $\texttt{elast3D}$ and $\texttt{dam3D}$, respectively. All of them share the same general structure. Given that Eqs.~\eqref{sol_u} and \eqref{sol_a} are linear, the subroutines consist of standard finite element procedures. Equations~\eqref{sol_u} and \eqref{sol_a} exhibit the general form:  

\begin{equation}
\vec{K}\vec{x} = \vec{b},
\end{equation}
\\
where $\vec{x}$ is a vector containing the nodal values of a primary variable. Therefore, the solution consists of determining the local coefficient matrix $\vec{K}^{el}$ and the local right hand side $\vec{b}^{el}$. Subsequently, the global coefficient matrix $\vec{K}$ and the global right hand side $\vec{b}$ are assembled, from where the solution is obtained after imposing boundary conditions. 

\subsubsection{Main module}\

The numerical setting results in an incremental staggered algorithm, as described in Algorithm~\ref{GeneralAlg}. For $n_{\mathrm{desp}}$ incremental displacements imposed on $\partial{\mathit{\Omega}}_u$, $[\vec{u}_{{n+1}}, \boldsymbol{\alpha}_{{n+1}}]$ are found iteratively using two independent tolerances $t_\mathrm{u}$ and $t_\mathrm{d}$, for a maximum number of iterations $k_{\mathrm{max}}$.

\begin{algorithm}
\centering
\begin{algorithmic}[1]

\FOR {$n=1 \ \ \textbf{to} \ \ n_{\mathrm{desp}}$}
\STATE Data input and initialization
\STATE k=0
\WHILE {\\ ($||\vec{u}^{k}_{{n+1}}-\vec{u}^{k-1}_{{n+1}}||_{2}>t_\mathrm{u} $ \ \ $\textbf{or} \ \ ||\boldsymbol{\alpha}^{k}_{{n+1}}-\boldsymbol{\alpha}^{k-1}_{{n+1}}||_{\infty}>t_\mathrm{d} $) \ \ $\textbf{and} \ \ k < k_{\mathrm{max}}$ } 
\STATE k=k+1
\STATE Obtain $\vec{u}_{{n+1}}^k$ from module \texttt{elast3D} \\
\STATE Obtain $\boldsymbol{\alpha}_{{n+1}}^k$ from module \texttt{dam3D}
\ENDWHILE
\ENDFOR
\end{algorithmic}
\caption{General routine}\label{GeneralAlg}
\end{algorithm}

\section{Numerical simulations}\

\begin{figure}
\centering
\includegraphics[scale=1.2]{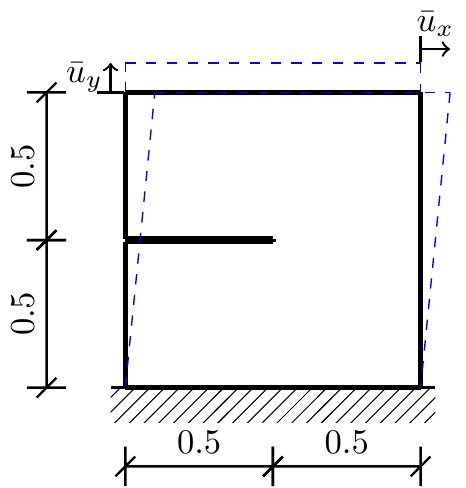}
\caption{\ General scheme for Experiment I and Experiment II.}
\label{setI}
\end{figure}

A square specimen with a horizontal notch in the middle is adopted for experiments involving both tension and shear loading. Monotonic displacements are imposed in the numerical experiments, as schematized in Fig.~\ref{setI}. For both tension and shear loading, the following parameters were adopted:

\begin{itemize}
\item $K = 121030$ MPa;
\item $\nu=0.227$;
\item $w_0 = 75.94$ MPa mm and
\item $\eta=0.052$ (MPa mm)$^{1/2}$ mm.
\end{itemize}

\paragraph{\textbf{Experiment I}}

The tension case is considered first, with $\bar{u}_x=0$ and $\bar{u}_y\geq0$ (Fig.~\ref{setI}). Vertical displacements are imposed on the top boundary from 0 to $\num{6e-3}$ mm, with increments of $\num{1e-4}$ mm, while the bottom boundary is fixed in both directions.

\begin{figure*}[!hbt]
\centering
\begin{subfigure}[b]{\textwidth}
\centering
\includegraphics[scale=0.16, trim={1.3cm 1cm 6.9cm 0.8cm},clip]{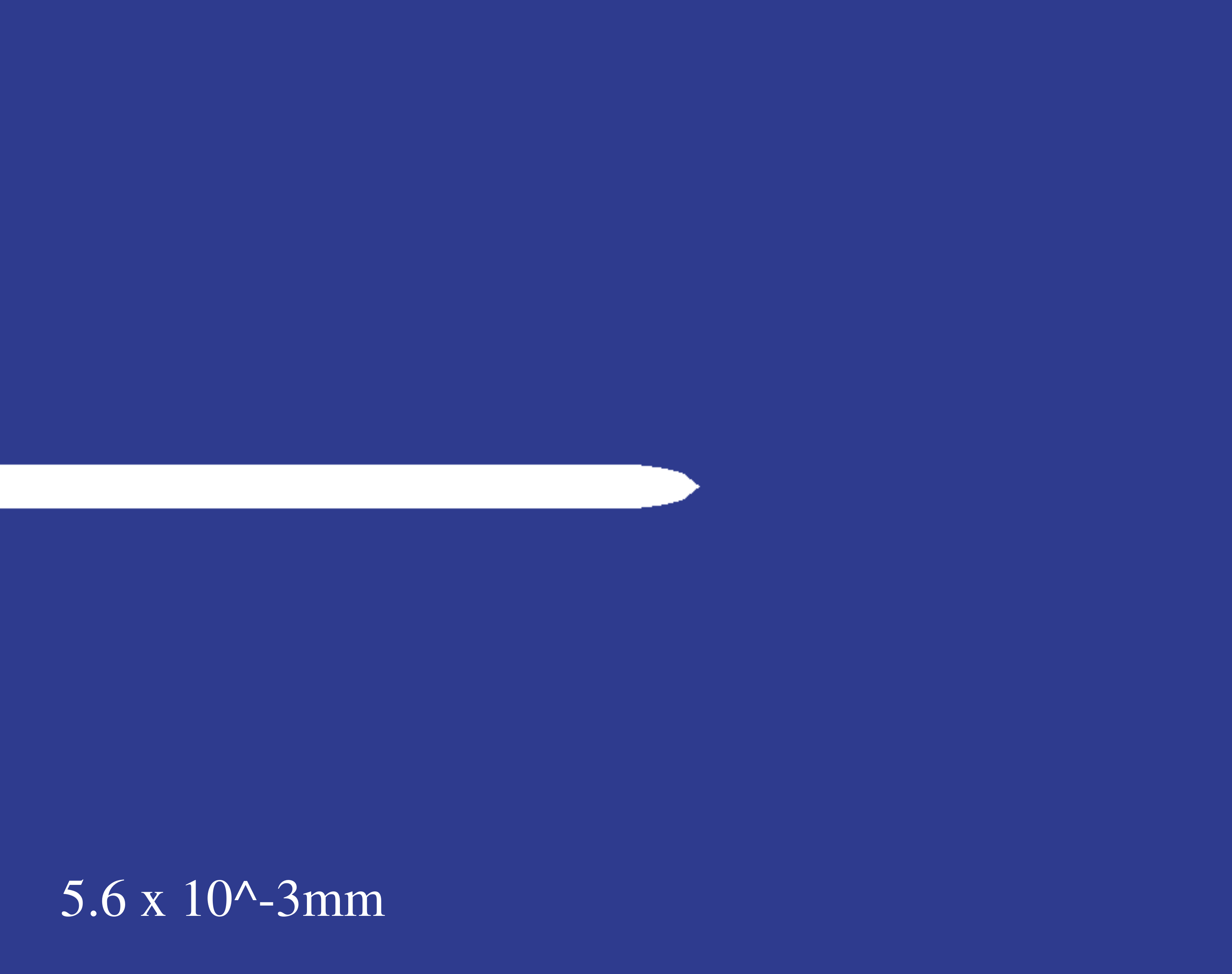} \ \ 
\includegraphics[scale=0.16, trim={1.3cm 1cm 6.9cm 0.8cm},clip]{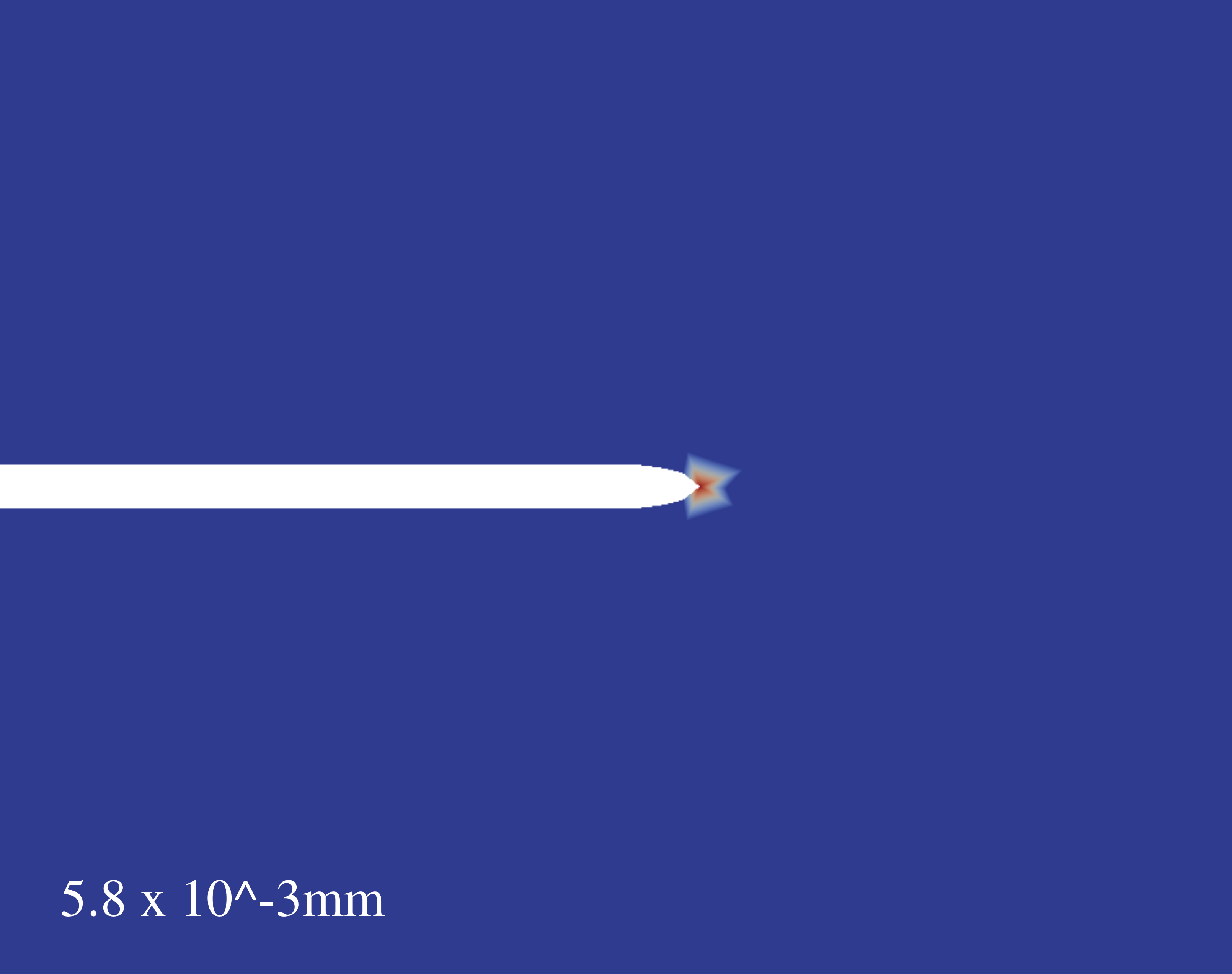} \ 
\includegraphics[scale=0.25, trim={8.5cm 1.4cm 0cm 1.7cm},clip]{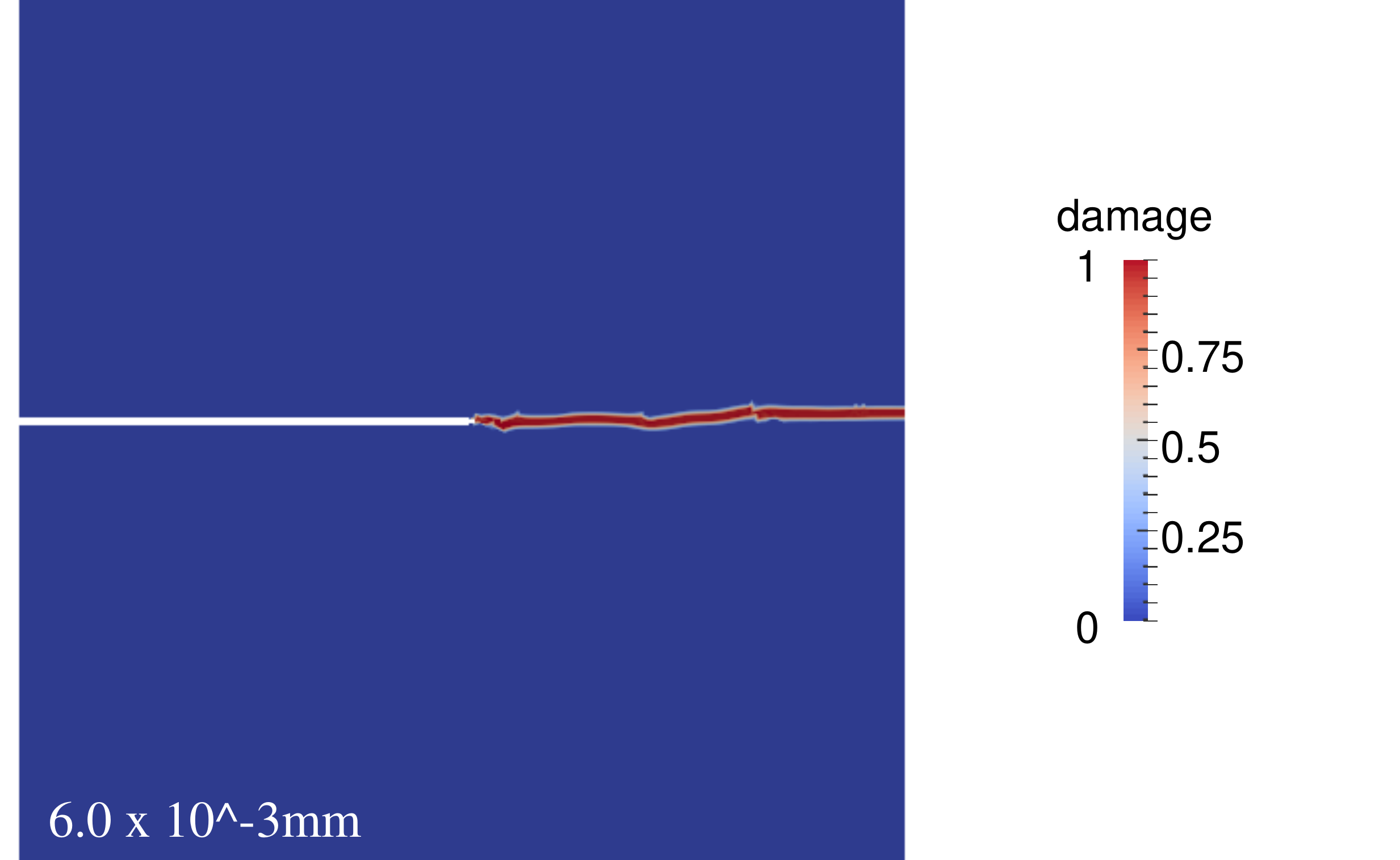}
\label{MieTDam}
\caption{\ Crack initiation close-up and crack propagation}
\end{subfigure}%
\hfill
\begin{subfigure}[b]{\textwidth}
\centering
\includegraphics[scale=0.25, trim={8.5cm 1cm 8.5cm 0cm},clip]{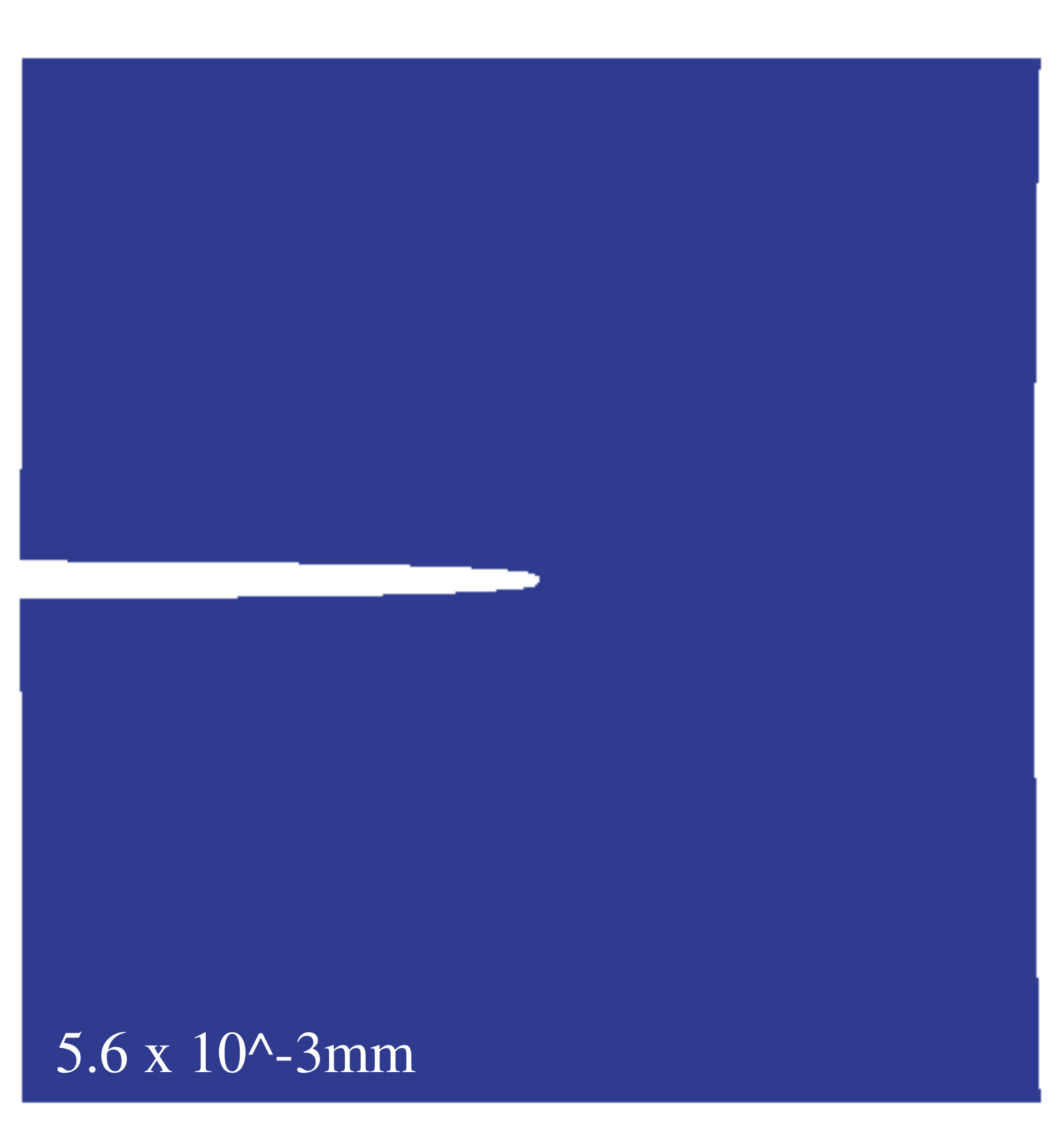}
\includegraphics[scale=0.25, trim={8.5cm 1cm 8.5cm 0cm},clip]{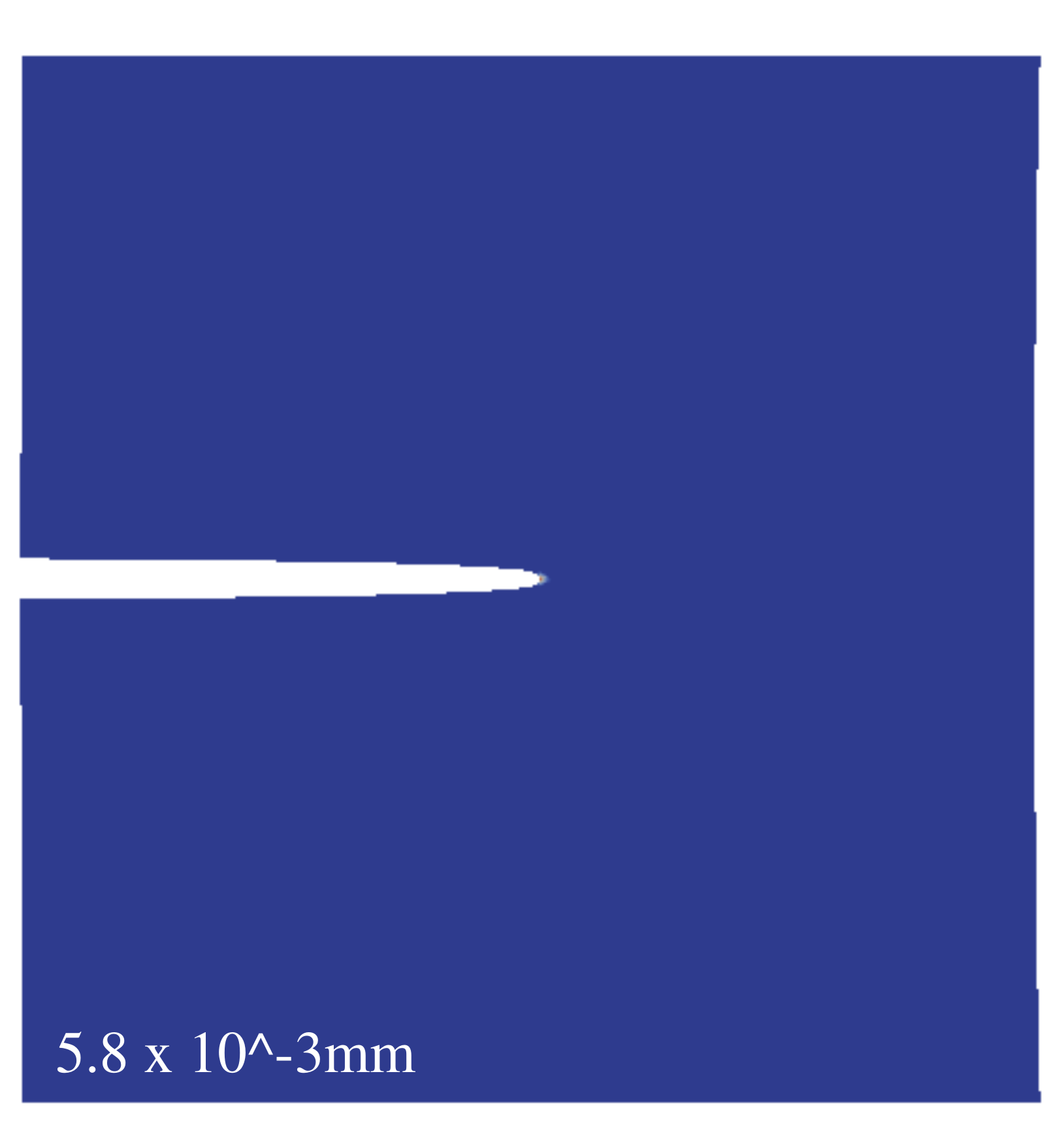}
\includegraphics[scale=0.25, trim={8.5cm 1cm 0cm 0cm},clip]{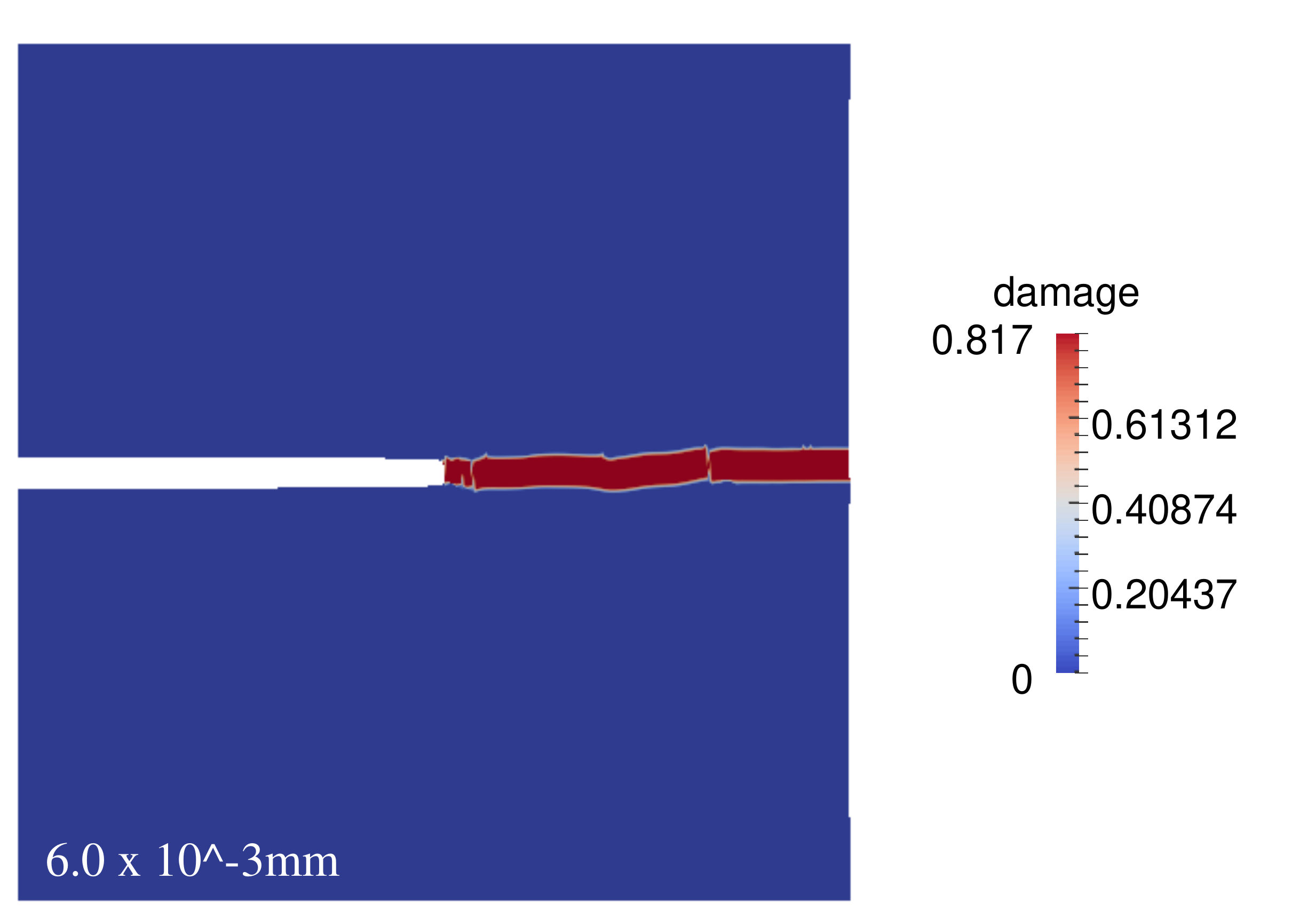}
\label{MieTDdef}
\caption{Deformed specimen}
\end{subfigure}%
\caption{\ Brittle crack evolution and deformed specimen for Experiment I: 2D simulation.}
\label{MieT}
\end{figure*}

\begin{figure}
\centering
\includegraphics[scale=0.6]{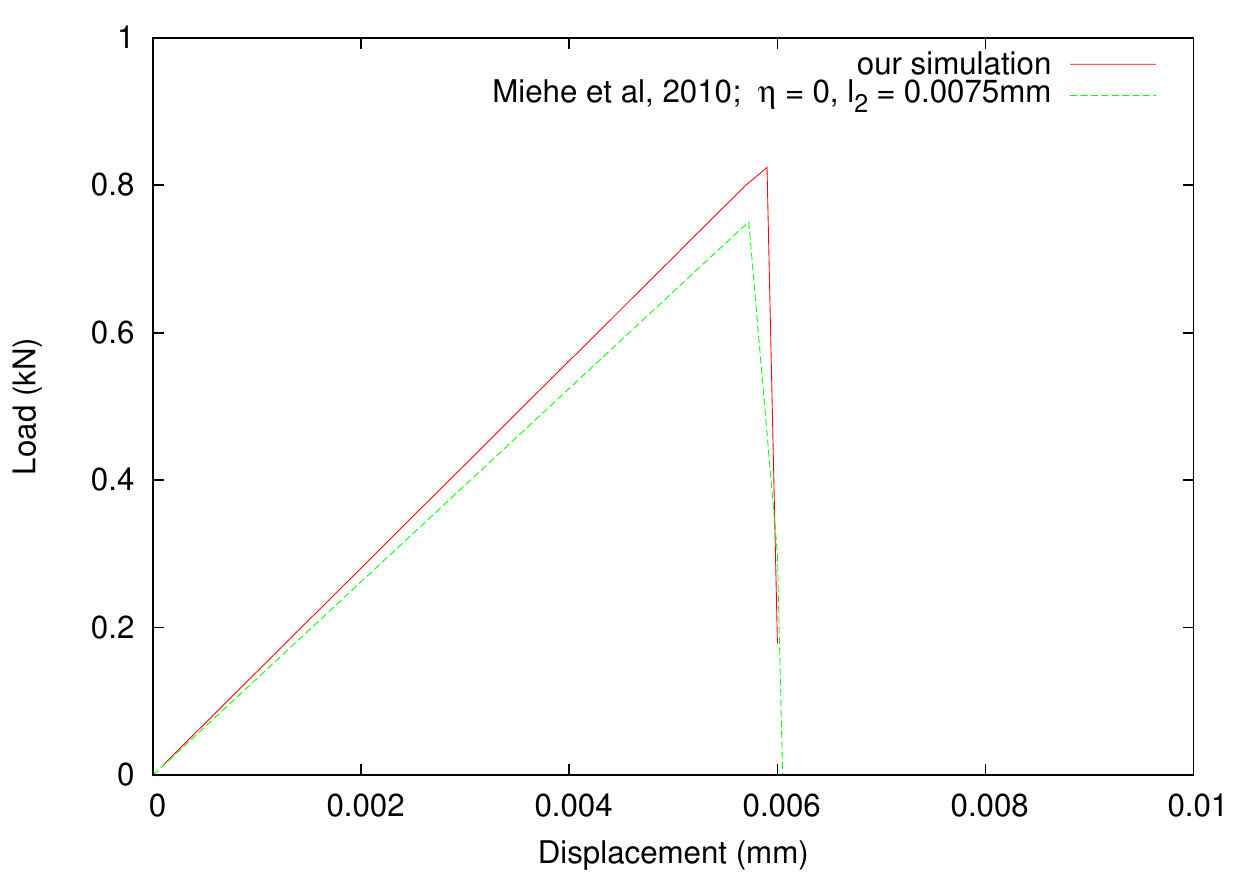}
\caption{\ Force-displacement curve for Experiment I: 2D simulation and comparison with the results of \citep{MieWelHof2010}.}
\label{fdI}
\end{figure}

The crack propagation and the deformed specimen are shown in Figure~\ref{MieT}. A single crack branch is initiated and propagates horizontally. As expected in brittle fracture, the specimen experiences a catastrophic-like failure mode, where the crack is initiated at the tip of the notch and propagates horizontally after a few load steps. This is reflected in the force-displacement curve shown in Figure~\ref{fdI}, where an abrupt drop in the load-carrying capacity is observed. The simulation was performed using 3401 bilinear quadrilateral elements. As shown in Figure~\ref{fdI}, the results closely resemble the experiment of \cite{MieHofWel2010}, where a finer mesh of 20000 linear triangles was used. 

\begin{figure}[hbt!]
\centering
\includegraphics[scale=0.26,trim={1.5cm 0cm 1.5cm 0cm},clip]{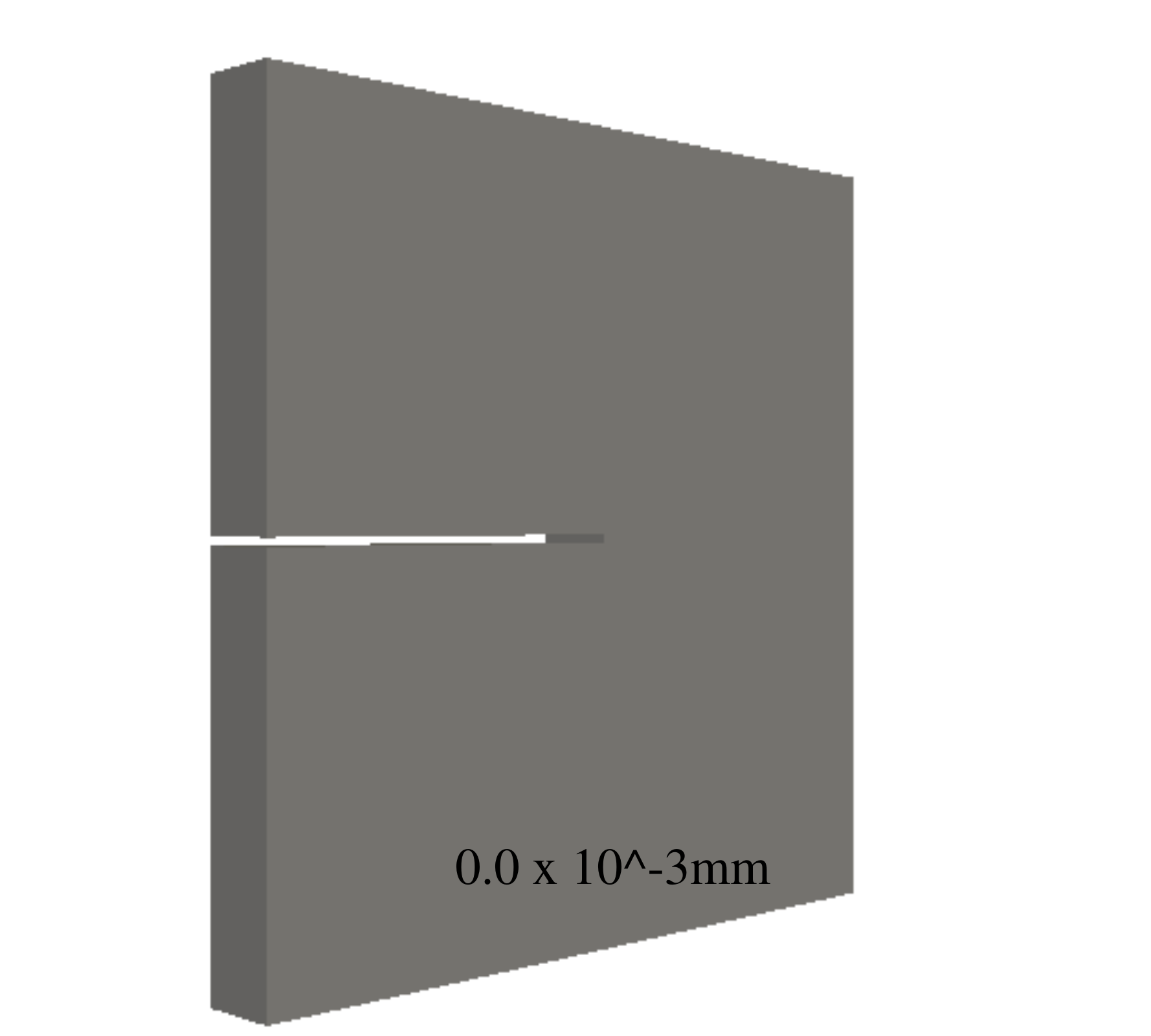}
\includegraphics[scale=0.26,trim={1.5cm 0cm 1.5cm 0cm},clip]{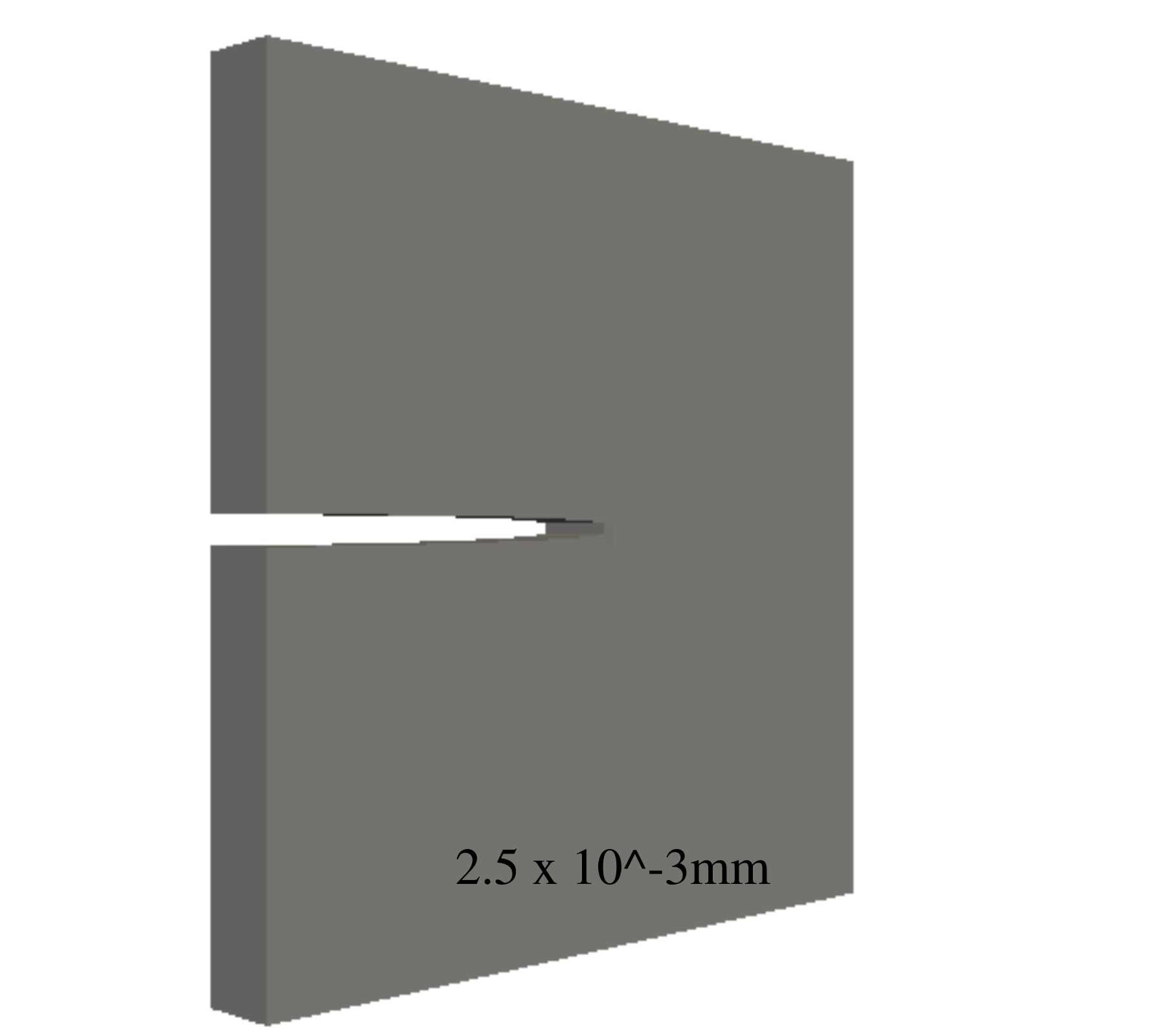}
\includegraphics[scale=0.26,trim={1.5cm 0cm 1.5cm 0cm},clip]{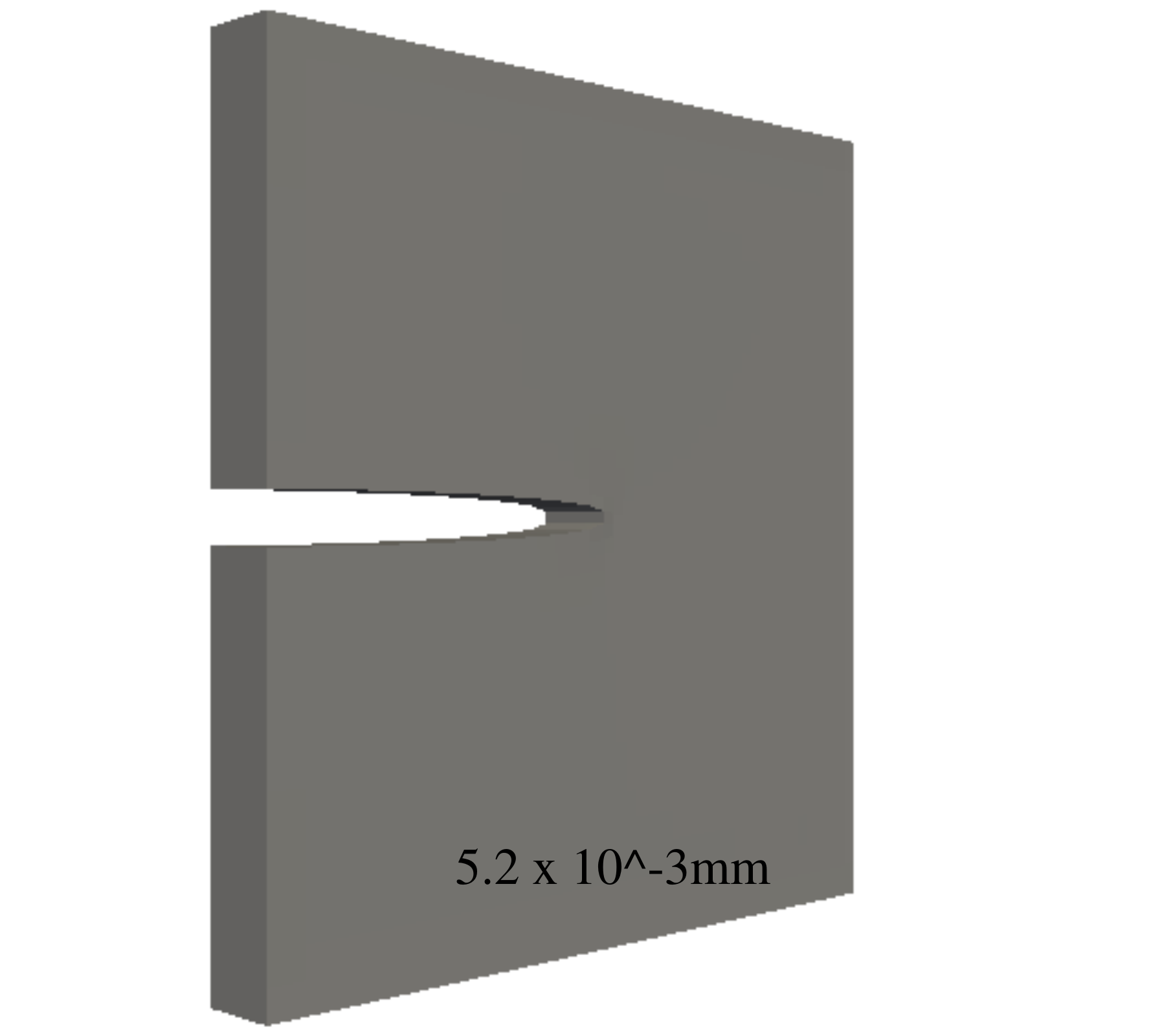}
\includegraphics[scale=0.26,trim={1.5cm 0cm 1.5cm 0cm},clip]{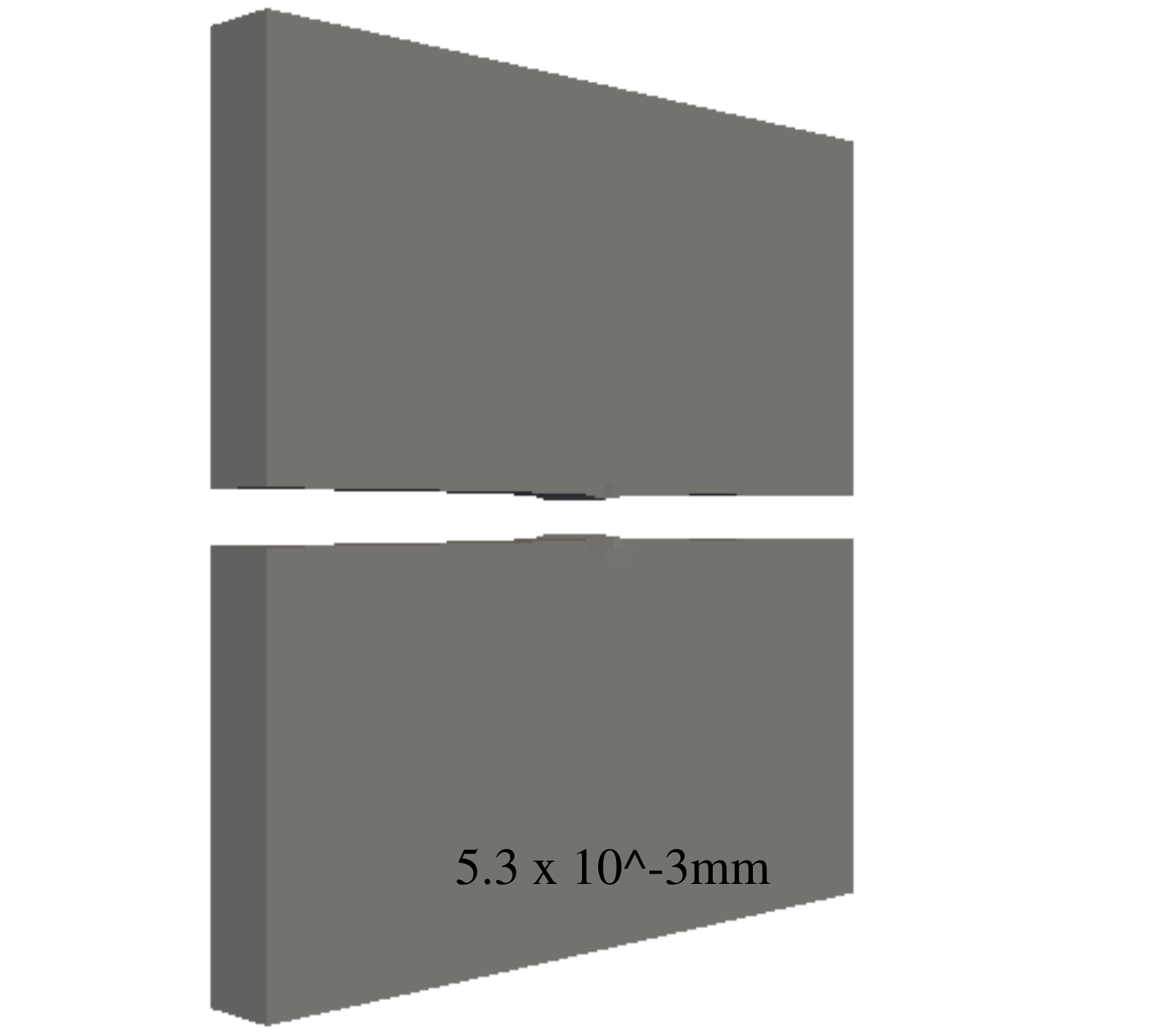}
\caption{\ Brittle crack for Experiment I: 3D simulation.}
\label{3D}
\end{figure}

\begin{figure}[hbt!]
\centering
\includegraphics[scale=0.6]{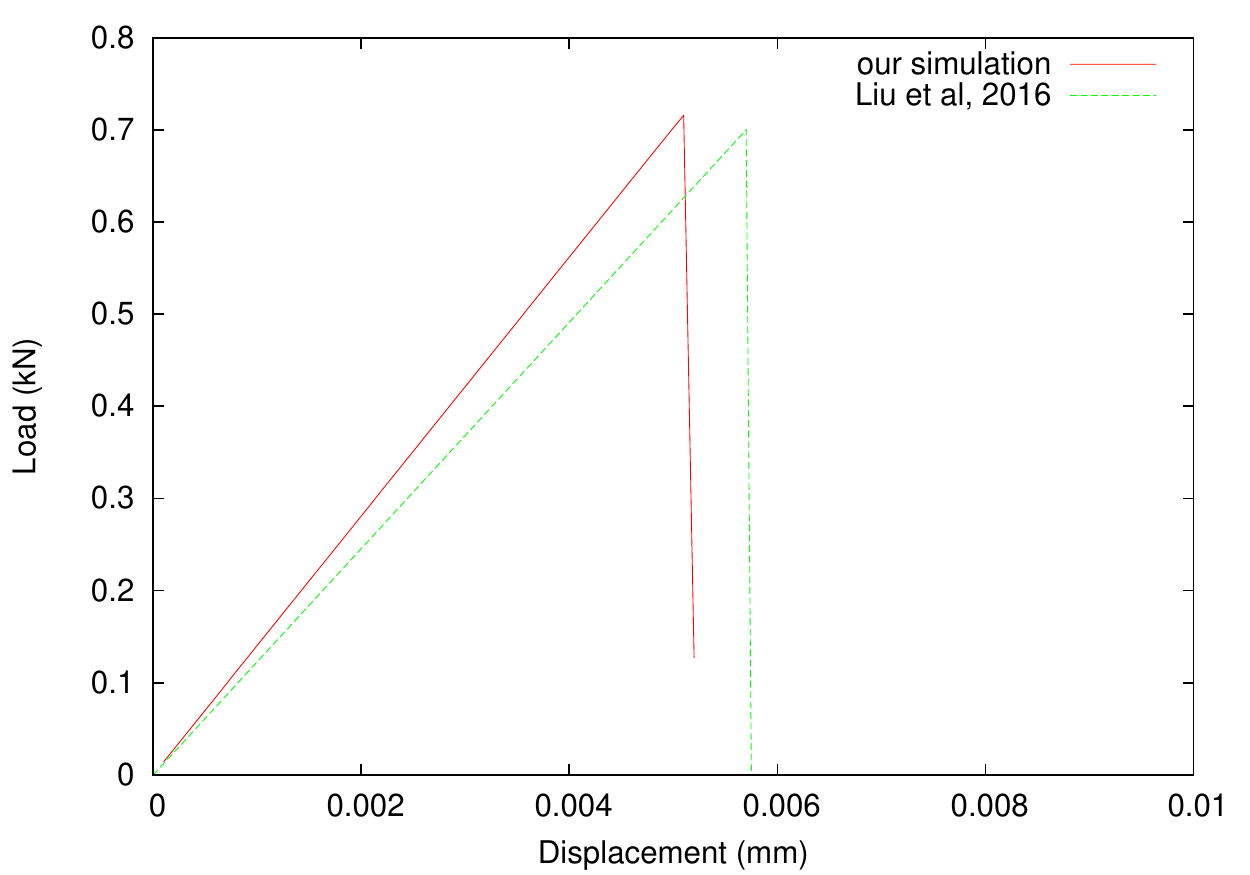}
\caption{\ Force-displacement curve for Experiment I: 3D simulation and comparison with the results of \cite{liu2016}.}
\label{3D2}
\end{figure}

For the same experiment, a three-dimensional (3D) simulation is shown in Figure~\ref{3D}. A uniform thickness of $0.1$ mm was used for the model, which was discretized with a mesh of 16500 hexahedral elements. The same displacements as in the two-dimensional simulation were imposed on the top boundary, while the bottom and lateral bottom boundary was fixed in all three directions and the lateral boundaries were fixed in the $z$ direction. In Fig.~\ref{3D}, the abrupt crack evolution can be observed better.  Figure~\ref{3D2} shows the resulting force-displacement curve, which is compared with the results obtained in \cite{liu2016}, where the implementation of the phase-field model without a damage threshold was performed in the commercial software Abaqus using 96896 hexahedral elements.

\paragraph{\textbf{Experiment II}}

We subject the same square specimen with a notch from Experiment I to shear loading, with $\bar{u}_x\geq0$ and $\bar{u}_y=0$ (Fig.~\ref{setI}). Incremental horizontal displacements are imposed on the top boundary, while the bottom boundary is fixed. Displacements are imposed from $0$ to $\num{12.7e-3}$ mm, with increments of $\num{1e-5}$ mm. The lateral boundaries are fixed in the vertical direction. For the simulation, a mesh of 3532 bilinear quadrilateral elements was used. 

\begin{figure*}[hbt!]
\centering
\begin{subfigure}[b]{\textwidth}
\centering
\includegraphics[scale=0.25, trim={8cm 1cm 9cm 1cm},clip]{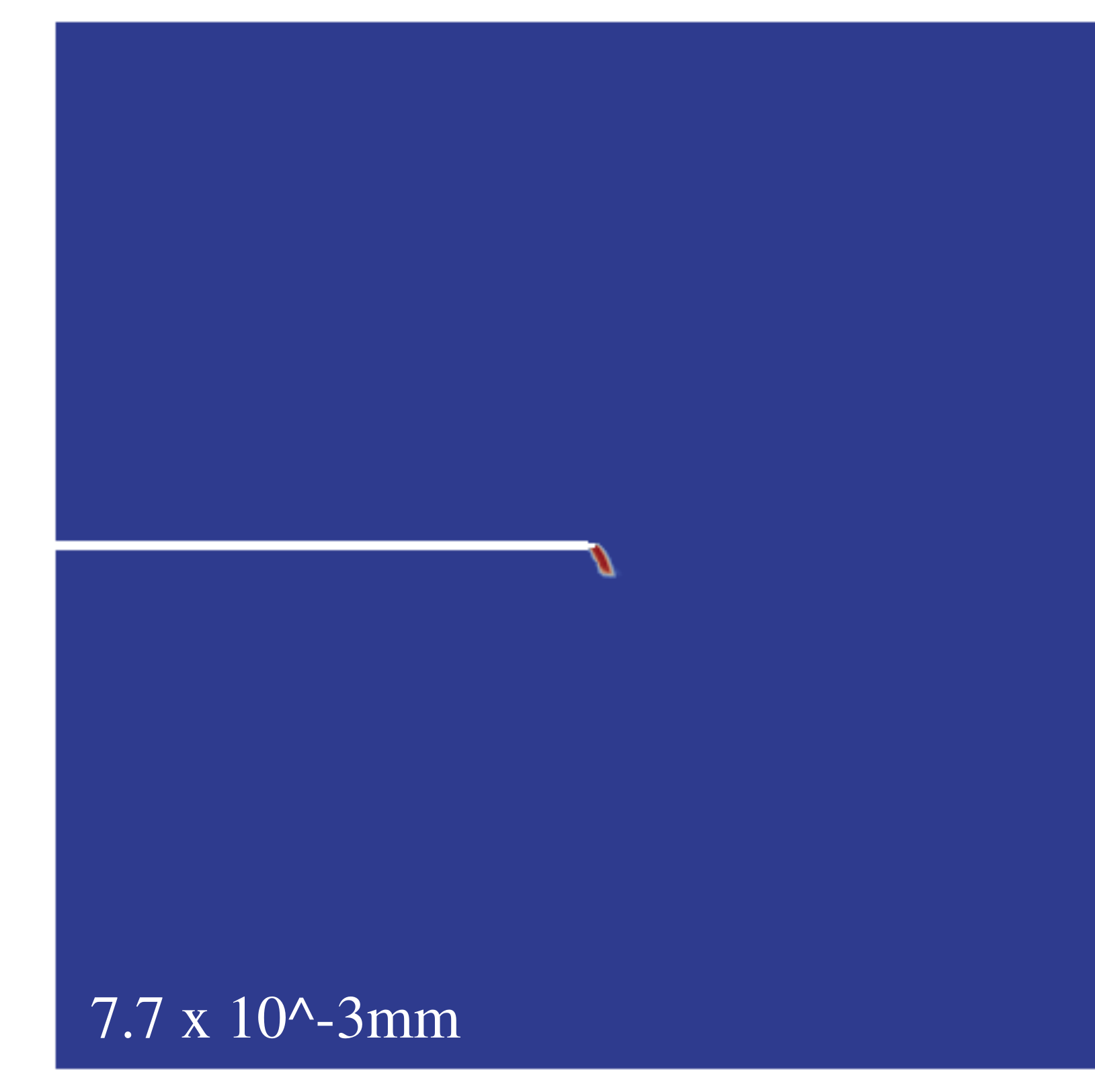}
\includegraphics[scale=0.25, trim={8cm 1cm 9cm 1cm},clip]{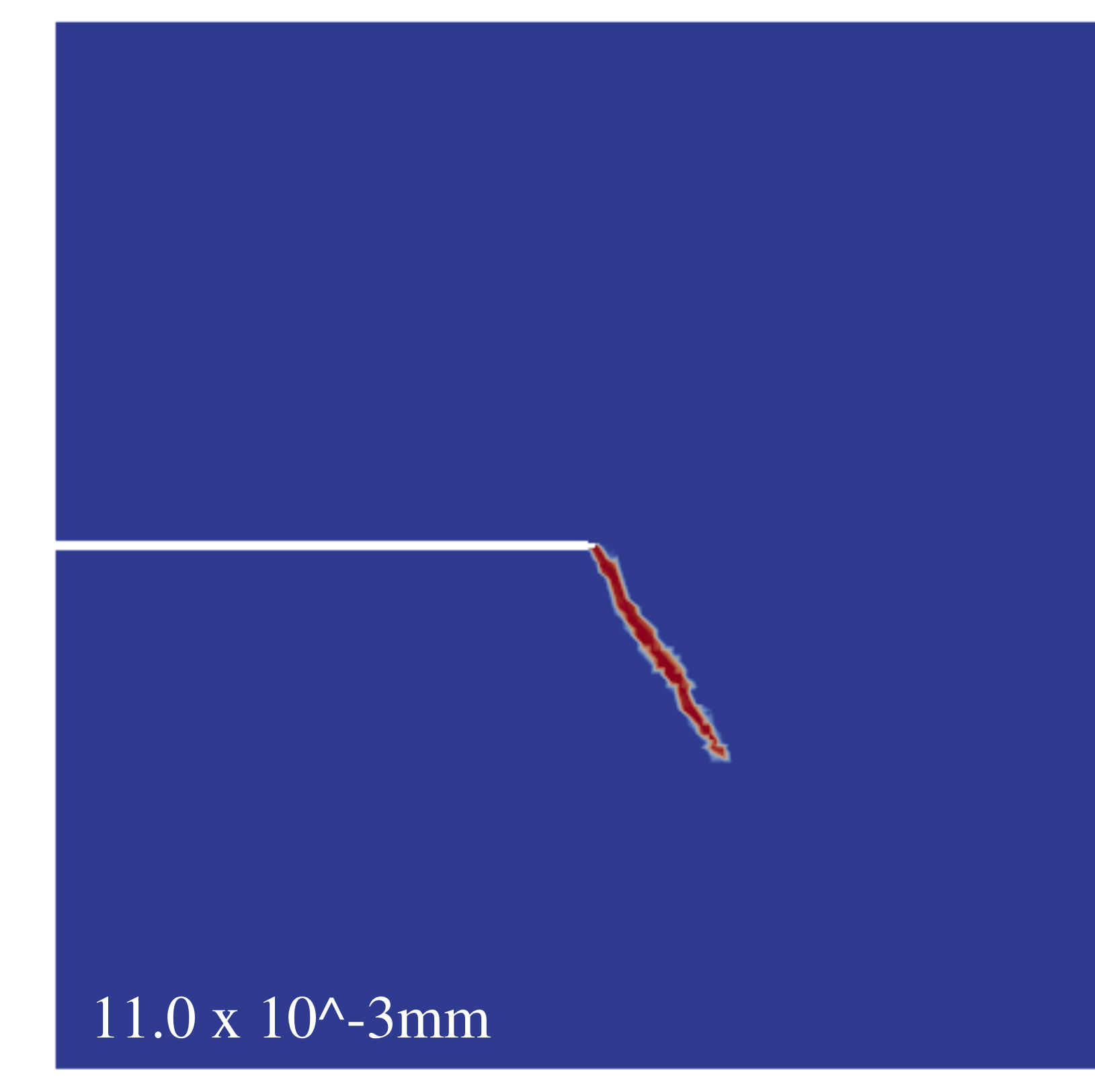}
\includegraphics[scale=0.25, trim={8cm 1cm 2cm 1cm},clip]{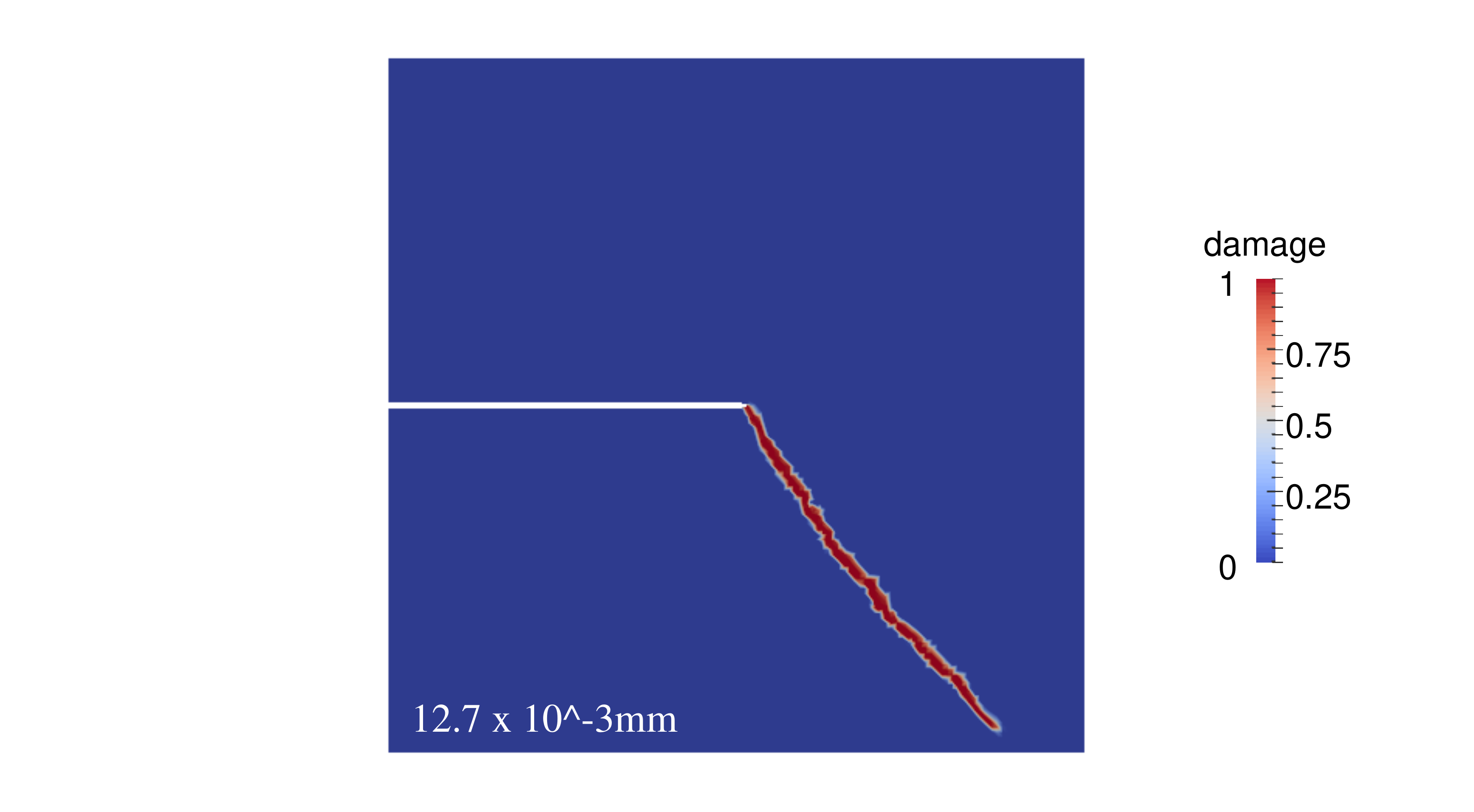}
\label{MieSDam}
\caption{Crack initiation and crack propagation}
\end{subfigure}%
\hfill
\begin{subfigure}[b]{\textwidth}
\centering
\includegraphics[scale=0.25, trim={8cm 1cm 9cm 0cm},clip]{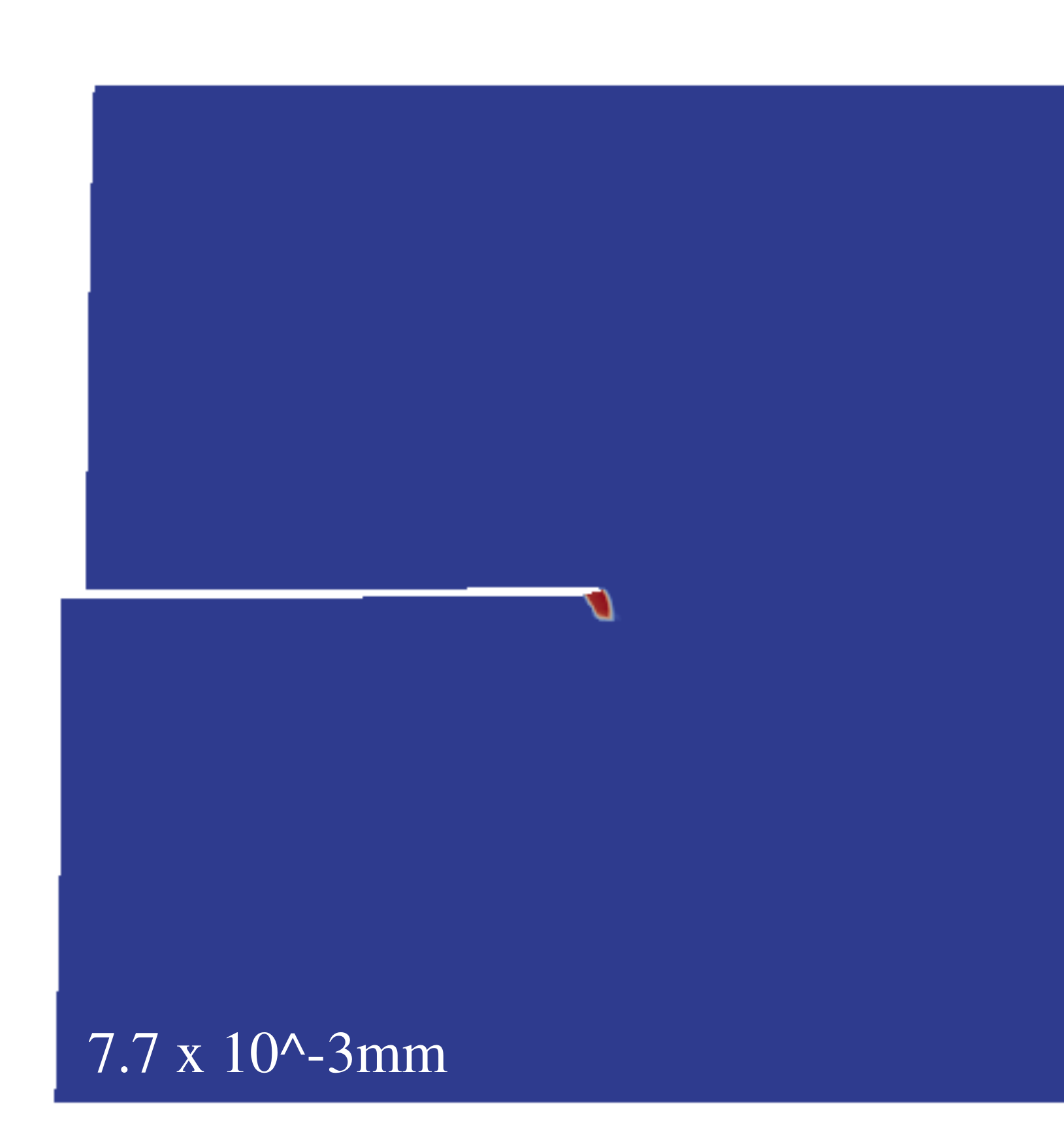}
\includegraphics[scale=0.25, trim={8cm 1cm 9cm 0cm},clip]{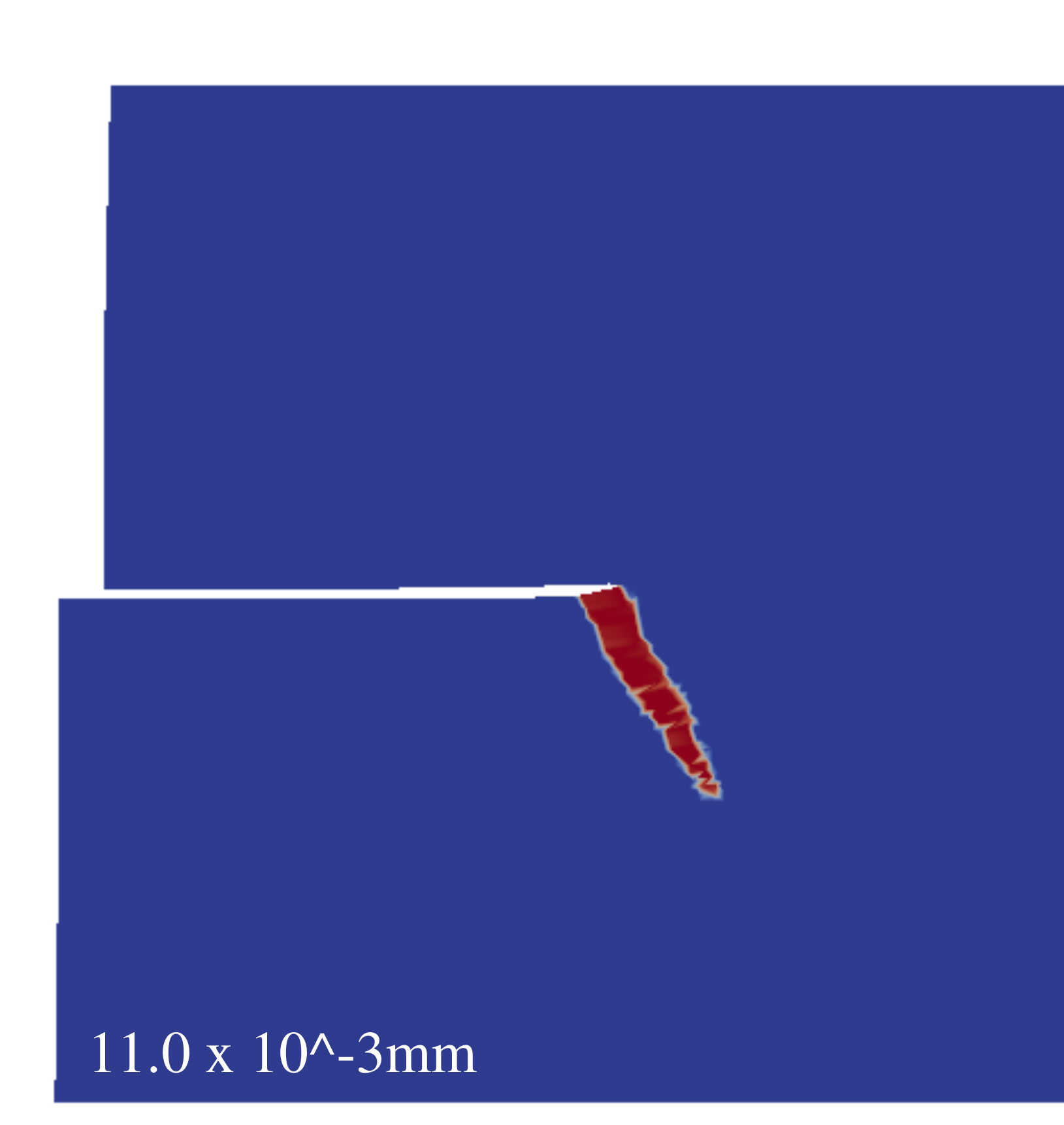}
\includegraphics[scale=0.25, trim={8cm 1cm 2cm 0cm},clip]{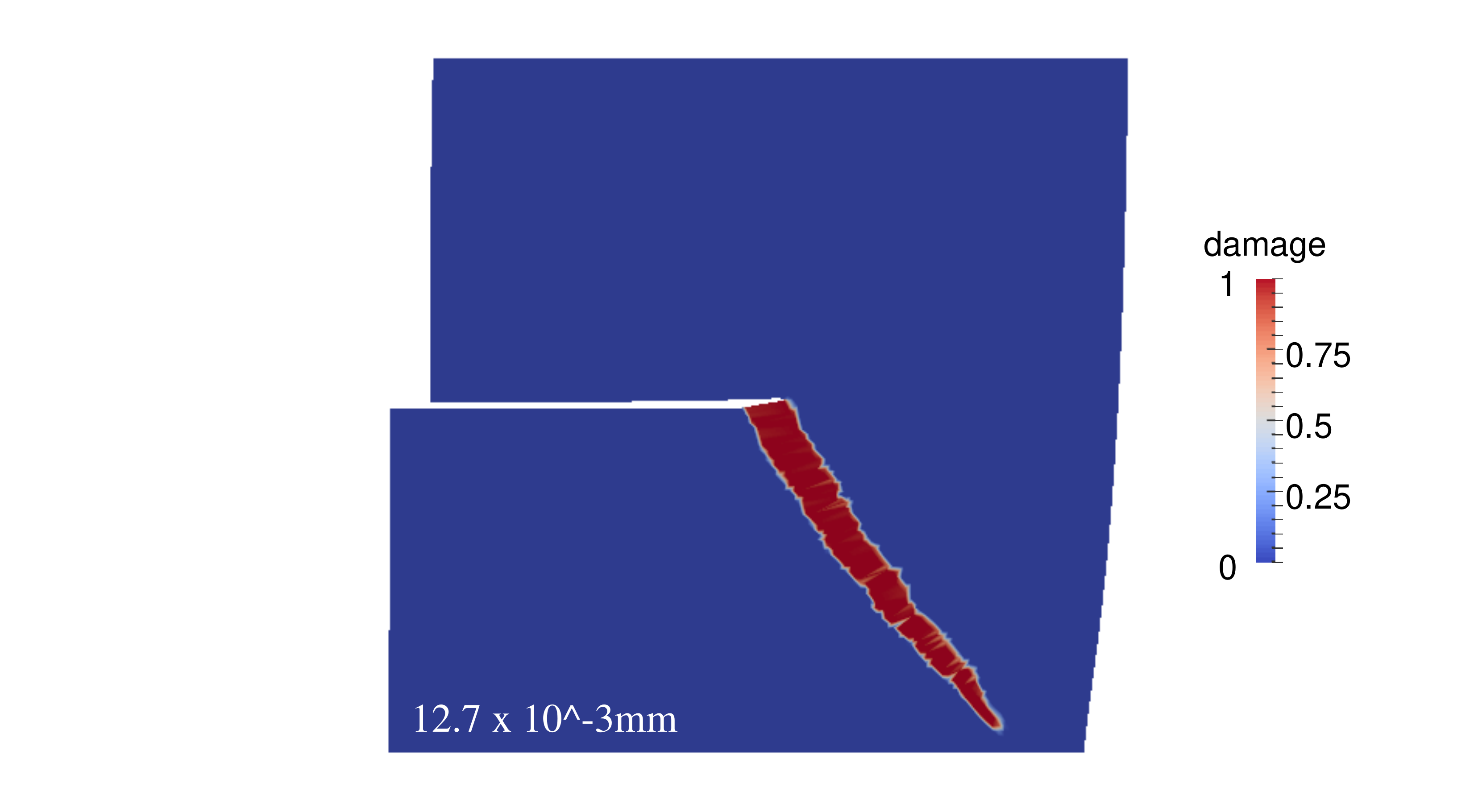}
\label{MieSDef}
\caption{Deformed specimen}
\end{subfigure}
\caption{\ Brittle crack evolution and deformed specimen for Experiment II.}
\label{MieS}
\end{figure*}

\begin{figure}[hbt!]
\centering
\includegraphics[scale=0.6]{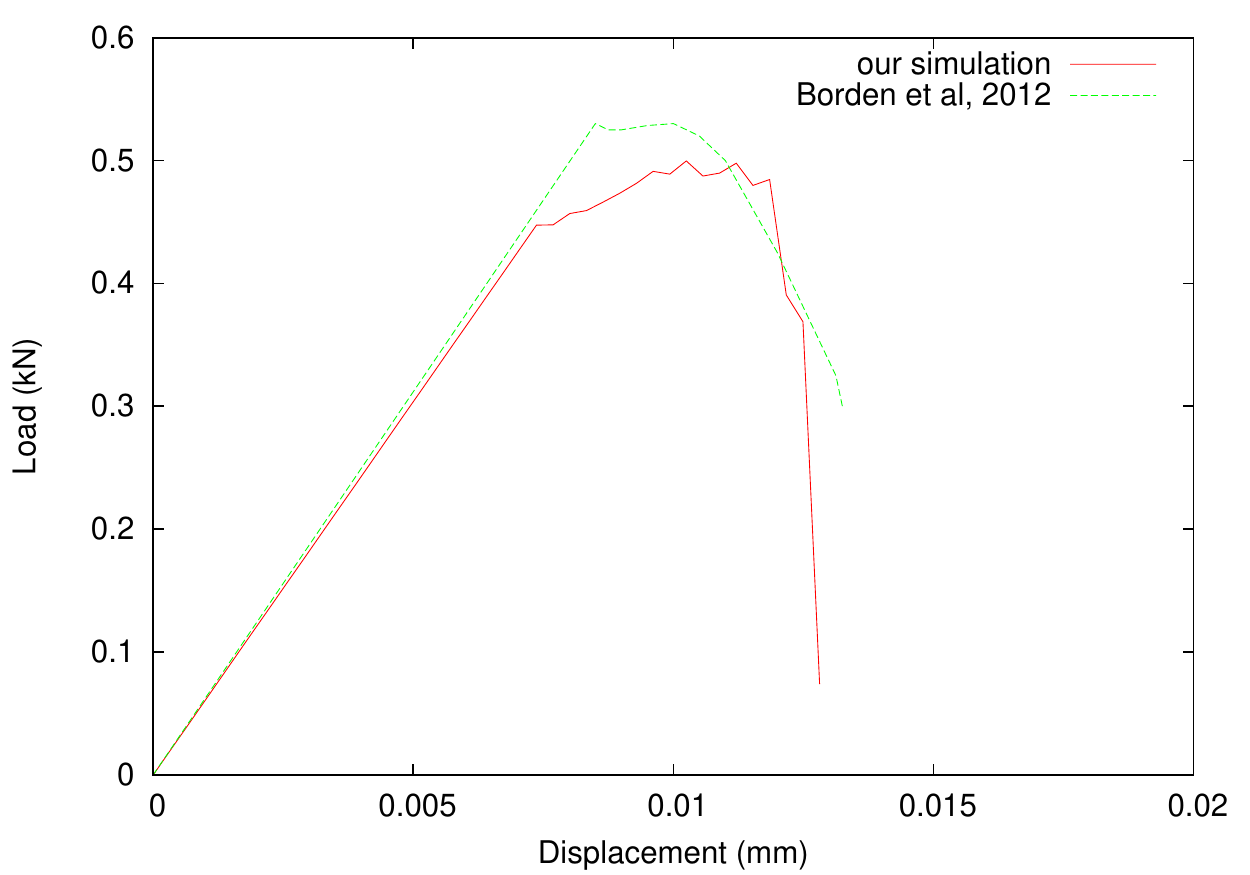}
\caption{\ Force-displacement curves for Experiment II: comparison with the results of \cite{Bord-Hugh2012}.}
\label{fdII}
\end{figure}

The crack propagation and the deformed specimen shown in Fig.~\ref{MieS} are a direct result of the decomposition into positive (due to tension) and negative (due to compression) energies in Eq.~\eqref{TensCompDecomp}. A single crack branch is initiated and propagates through regions of intense positive stress. Along with the force-displacement curve shown in Fig.~\ref{fdII}, these results resemble the results of \cite{MieHofWel2010} and \cite{Bord-Hugh2012}. The former used 30000 linear triangles, while the latter applied cubic T-splines with 5587 cubic basis functions.

%
%
%

\section{Conclusions}

We herein presented a multidimensional model for the description of brittle fracture using the phase-field approach, which was written based on a consistent energetic formulation. As opposed to the most popular approaches of phase-field fracture, we adopted constitutive assumptions in the spirit of gradient-enhanced damage models, which allow for a damage threshold. We consider this to be an attractive feature because an elastic stage is included in the deformation process. The simulations presented indicate good agreement with the results obtained in the literature using the model without an elastic stage. 

One of the main features of the formulation explored in this work is its variational character. This allows for an elegant approach, resulting in a powerful tool to describe the behavior of solids undergoing brittle fracture. Moreover, the resulting numerical problem can be solved in a relatively straightforward manner. We have explicitly presented a simple 3D finite element formulation that could be easily implemented in standard finite element routines. Although the staggered approach is known to result in slow convergence when compared to monolithic solution schemes, no major convergence issues were observed in the performed simulations other than the expected increase in iterations at the onset of the softening regime. 

The obvious extension of this study is the introduction of ductile behavior, which is the subject of another article. In addition, since high run times were experienced, particularly for 3D simulations, efficient numerical implementations are crucial for further developments, which is a topic of further study.

\paragraph{\textbf{Acknowledgements}}

We gratefully acknowledge the financial support of CYTED (Ibero- American Program to Promote Science and Technology) through the CADING network (Ibero-American Network for High Performance Computing in Engineering). The support of the EU H2020-MSCA-RISE-2016 project ``BESTOFRAC: Environmentally best practices and optimisation in hydraulic fracturing for shale gas/oil development'' is also acknowledged.


\end{document}